\documentclass[10pt,longbibliography,physrev,aps,twocolumn]{revtex4-2}
\usepackage{graphicx}
\usepackage{xcolor}
\usepackage{braket}
\usepackage{amsmath,bm}
\usepackage{enumitem}
\usepackage{mathrsfs} 
\usepackage{amssymb}
\usepackage[colorlinks]{hyperref}

\DeclareMathOperator{\tr}{tr}

\DeclareMathOperator{\arccot}{arccot}
\DeclareMathOperator{\rnk}{rnk}

\begin{document}

\title{Fragmented exceptional points and their bulk and edge realizations in lattice models}

\author{Subhajyoti Bid}
\affiliation{Department of Physics, Lancaster University, Lancaster, LA1 4YB, United Kingdom}
\author{Henning Schomerus}
\affiliation{Department of Physics, Lancaster University, Lancaster, LA1 4YB, United Kingdom}
\date{\today}
\begin{abstract}
Exceptional points (EPs) are spectral defects displayed by non-Hermitian systems in which multiple degenerate eigenvalues share a single eigenvector. 
This distinctive feature makes systems exhibiting EPs
more sensitive to external perturbations than their Hermitian counterparts, where degeneracies are nondefective diabolic points. 
In contrast to these widely studied cases, more complex non-Hermitian degeneracies in which the eigenvectors are only partially degenerate are poorly understood. Here, we characterize these  fragmented exceptional points (FEPs) 
systematically from a physical perspective, 
and demonstrate how they can be induced into the bulk and edge spectrum of two-dimensional and three-dimensional lattice models, exemplified by non-Hermitian versions of a Lieb lattice and a higher-order topological Dirac semimetal. The design of the systems is facilitated by an efficient algebraic approach within which we provide precise conditions for FEPs that can be evaluated directly from a given model Hamiltonian. The free design of FEPs significantly opens up a new frontier for non-Hermitian physics and expands the scope for designing systems with unconventional response characteristics.
\end{abstract}

\maketitle
\section{Introduction}
In the past decade, topological phases in Hermitian systems, including topological insulators and superconductors \cite{RevModPhys.83.1057,RevModPhys.82.3045,bernevig,toposuperbeena,asboth2016,Sato_2017} as well as Dirac and Weyl semimetals  \cite{Murakami2007,Burkov2016,armitage2018weyl,weylannual} have drawn substantial interest 
due to a wide range of possible applications.
One of the key features of these topological materials is the existence of gapless edge and surface states that are  protected by the underlying symmetries of the bulk Hamiltonian. Due to this protection, these states are immune to symmetry-preserving disorder present in the system.  
The investigation of such topological phases significantly expanded with the identification of higher-order topological insulators with quantized quadruple moments 
\cite{multipole1,multipole2,HOTI1,HOTI3}, including 3D higher-order Dirac semimetals (HODSM) \cite{Hodsm,Wieder2020} and 3D higher-order Weyl semimetals \cite{Howsm,howsm2,howsm3}. Unlike conventional topological insulators that display edge or surface states at their boundaries, these materials feature topological corner or hinge states when they are confined in two directions.

With the recent resurgence and expansion of non-Hermitian  physics \cite{Hatano1996,moiseyev,Bender, cao2015dielectric,El-Ganainy2018,Ashida}, the study of topological materials has rapidly extended to their non-Hermitian equivalents, see for example Refs.~\cite{Schomerus13,Malzard:2015,nhtopo,kawabata2019symmetry,HOTI2,nhdirac,Denner2021,bidnh,nhtopo3}. In the non-Hermitian setting, the band structure and energies of topologically protected states generally become complex, and the associated eigenstates are no longer constrained to be mutually orthogonal.
One of the main resulting features is the emergence of distinct branch points in the complex energy eigenspectrum, known as exceptional points (EPs),
where both the eigenenergies and eigenvectors coalesce  \cite{kato,Dembowski2001,Heiss_2004,Heiss_2012,Berry2004}. 
This behavior is distinct from that at diabolic points (DPs), the conventional degeneracies in Hermitian systems, where only the eigenenergies merge. As topological transitions of the band structure and the emergence of topologically protected states are linked to the closing of band gaps, and the bands and protected states are themselves now located in the complex energy plane, EPs significantly expand the scope of topological phases, and this is further enriched by an expanded set of possible symmetries \cite{kawabata2019symmetry}.

Exceptional points are a universal phenomenon in any non-Hermitian system. In the parameter space of the system, the locations of the EPs form manifolds, known as exceptional surfaces \cite{exceptional_lines1,exceptional_lines2,exceptional_lines3,exceptional_lines4}, whose structure depends on the complexity of the EP, as well as constraints imposed by symmetries. 
These features result in
directly observable  phenomena, such as a nonlinear power-law response to parametric perturbations  \cite{jansensor,Chen2017,Lai2019,Wiersig2020} and a super-Lorentzian response to quantum noise and external driving \cite{Yoo:2011,Schomerus2020,Hashemi2022}.
In both response settings, the qualitative and quantitative signatures \cite{jan1,jan2,Schomerus2024,bid2024uniform} depend on the order $n$ of the EP, where an $n$th-order EP (short, an EP$n$)  is obtained by merging $n$ eigenvalues, with the complete collapse of the eigenstates into a single one \cite{Demange_2012,ryu1,hoep,ipsita}. EP$n$s have already been realized in various experiments demonstrating a significant increase in the system's sensitivity \cite{Hodaei2017,Wu:21,Yang2023,Wiersig:20}. 

Non-Hermitian degeneracies with only partially degenerate eigenstates
have received much less attention even though they are as ubiquitous as higher-order EPs.  
Mathematically, these most general degeneracies can be classified by partial multiplicities that describe the fragmentation of the generalized eigenspace into the subspaces associated with the surviving eigenstates \cite{hormander1994}. 
Just as higher-order EPs, these \textit{fragmented exceptional points} (FEPs) naturally appear in specific positions on the exceptional surface \cite{kunst,kunst2,koenig2024,Ryu2024}. 
However, their appearance poses a significant challenge for the description and classification of non-Hermitian systems, as the standard perturbative approach, Arnold-Jordan normal form theory \cite{Arnold1971}, breaks down for these scenarios.
This state of affairs also hampers the systematic description and design of systems and the identification of the physical manifestations of FEPs.
Therefore, obtaining a systematic handle on FEPs is a critical step to
complete the description and understanding of non-Hermitian systems both conceptually as well as physically, and realize the full scope of their phenomenology.

Motivated to overcome this roadblock, we establish in this work a systematic and efficient framework to design, classify, and characterize systems exhibiting FEPs.
Furthermore, we apply this framework to address the interplay of non-Hermiticity and symmetry in specific two-dimensional and three-dimensional lattice models, and formulate exact analytical conditions for the formation of FEPs in bulk and boundary spectra.
The framework directly links to the qualitatively different response of FEPs to perturbations and driving, 
and freely applies to any model based on an effective non-Hermitian Hamiltonian matrix. This delivers systematic algebraic conditions in terms of the so-called modes of the adjugate matrix, which can be directly evaluated for any given Hamiltonian matrix and yield explicit conditions in terms of microscopic model data. Given the physical starting point, the very same quantities directly determine the physical response of the system to external and parametric perturbations. 
For the models, we first consider the two-dimensional Lieb lattice and apply the framework to construct and classify non-Hermitian versions that host diverse  FEPs with different degeneracy structures in the bulk dispersion. 
As a second model, we turn to a three-dimensional HODSM that supports topologically protected hinge states, and design variants that display FEPs in the bulk and boundary spectra. The FEP formation for hinge states reveals a highly intricate interplay of nonreciprocal non-Hermitian couplings and symmetry constraints, illuminating the richness of mechanisms determining FEPs. 
The general framework and specific applications presented here provide a template to construct and classify models with unconventional spectral and response features across a wide range of platforms, significantly expanding the scope to understand and design non-Hermitian physical systems and devices.

The outline of the paper is as follows. After
briefly recapitulating the role of DPs and EPs in the spectral analysis of Hermitian and non-Hermitian systems in  Sec.~\ref{sec:background}, we provide in Sec.~\ref{sec:peps} our efficient and systematic framework for a mathematical and physical description of FEPs.
In Sec.~\ref{sec:NH-Lieb} we apply this formalism to investigate the two-dimensional Lieb lattice, and use it to construct and classify non-Hermitian versions displaying FEPs of all admissible types in the bulk dispersion relation.
In Sec.~\ref{sec:NH-HODSM} we extend these consideration to a three-dimensional HODSM and demonstrate the existence of diverse FEPs in the bulk eigenspectra of appropriately designed non-Hermitian variants. In Sec.~\ref{sec:hinges} we discuss the hinge states in these HODSM variants, resulting in the emergence of fragmented exceptional lines with different degeneracy structures. We give our conclusions in Sec.~\ref{sec:conclusion}. A key technical step in the derivation of the employed formalism is provided in  Appendix \ref{app:rankcondition}. 

\section{Exceptional and diabolic points}
\label{sec:background}
To prepare our discussion of FEPs,  we briefly review the notion of an EP of order (EP$n$), which is the most commonly studied spectral degeneracy scenario of non-Hermitian physics, and contrast it to the notion of an $n$-bolic point, such as a diabolic point, which constitutes the spectral degeneracies of Hermitian physics.

We place our results in the context of systems with an effective non-Hermitian Hamiltonian, represented by an $N\times N$ dimensional square matrix matrix $H$. The eigenvalues are determined from the condition
\begin{equation}
p(E)\equiv\mathrm{det}\,(E\openone -H)=0,
\end{equation}
hence, are given by the roots of a polynomial $p(E)$ of order $N$. 
The algebraic multiplicity $\alpha_i$ of an eigenvalue $E_i$ is the order of the root, and these multiplicities add up to $\sum_i\alpha_i=N$.

Given an eigenvalue $E_i$, the geometric multiplicity $\gamma_i$ determines 
the number of linearly independent right eigenvectors $\mathbf{u}_{i,j}$, $j=1,2,\ldots,\gamma_i$, fulfilling
\begin{equation}
\label{eq:eigenvalueproblem}
(E_i\openone -H)\mathbf{u}_{i,j}=0,
\end{equation} 
as well as 
the number of linearly independent left eigenvectors $\mathbf{v}_{i,j}$, $j=1,2,\ldots,\gamma_i$, fulfilling
\begin{equation}
\label{eq:eigenvalueproblem2}
\mathbf{v}_{i,j}(E_i\openone -H)=0.
\end{equation} 
The geometric multiplicity  is therefore determined by the rank of these homogeneous systems of equations,
\begin{equation}
\gamma_i=N -\mathrm{rnk}\,(E_i\openone - H),
\label{eq:gm}
\end{equation}
and is constrained according to
\begin{equation}
    1\leq \gamma_i \leq \alpha_i.
\end{equation}

For a nondefective system, the multiplicities $\alpha_i=\gamma_i$ coincide for all eigenvalues. Hence, the number of linearly independent right (or left) eigenvectors $\sum_i\gamma_i=N$, so that each set of eigenvectors forms a basis. 
The system can then be diagonalized by a suitable similarity transformation 
\begin{equation}
\label{eq:diag}
\Lambda=VHU,
\end{equation}
where $\Lambda$ is a diagonal matrix containing the eigenvalues $E_i$ of $H$ according to their multiplicity, $U$ contains the corresponding right eigenvectors $\mathbf{u}_{i,j}$ as its columns, and $V$ contains the corresponding left eigenvectors $\mathbf{v}_{i,j}$ as its rows.
Given that for a similarity transformation $V=U^{-1}$, the
eigenvectors constructed in this way fulfill the biorthogonality condition
\begin{equation}
\mathbf{v}_{i,j}\mathbf{u}_{i',j'}=\delta_{ii'}\delta_{jj'}.
\end{equation}

Degenerate eigenvalues with $\alpha_i=\gamma_i=2$ are known as diabolic points (DPs), and higher order degeneracies where $\alpha_i=\gamma_i=n$ with some $n>2$ can be referred to as $n$-bolic points. Analogously, maximally defective eigenvalues with $\gamma_i=1$ but $\alpha_i=n$ with some $n>1$ are known as an $n^{th}$-order exceptional points (EP$n$s). At these EPs,  $n$ algebraically degenerate eigenvalues share a single, unique, eigenvector. 
The existence of EPs therefore renders a system nondiagonalizable, leading to the distinct physical signatures in the response to driving and perturbations surveyed in the introduction. 

\section{Fragmented exceptional points\label{sec:peps}}

In between the $n$-bolic point and the EP$n$, a degenerate eigenvalue $E_i$ can have several linearly independent eigenvectors, corresponding to FEPs with a geometric multiplicity $\alpha_i>\gamma_i>1$. 
We first detail the mathematical and physical signatures of these most general spectral scenarios,  and then provide an efficient algebraic framework by which we can determine these signatures directly from model data.
\subsection{Partial multiplicities\label{sec:partialmult}}
 
For degenerate eigenvalues with $\alpha_i>\gamma_i>1$ one can define partial multiplicities
$ l_{i,1} \geq l_{i,2} \geq l_{i,3} \dots \geq l_{i,\gamma_i}\geq 1$ 
that designate how many algebraically degenerate eigenvalues are associated with a given eigenvector. These partial multiplicities therefore partition the algebraic multiplicity according to 
\begin{equation}
    l_{i,1} + l_{i,2} + l_{i,3}  + \dots + l_{i,\gamma_i} = \alpha_i,
    \label{partial}
\end{equation}
where each partition of $\alpha_i$ describes a different spectral scenario. Furthermore, by including the cases of a single partial multiplicity $l_{i,1}=\alpha_i$ (EP$n$ with $n=\alpha_i$) and the case of all $l_{i,1}=l_{i,2}  = \dots = l_{i,\gamma_i}=1$ ($n$-bolic point with $n=\gamma_i$), the specification of the partial multiplicities covers all possible degeneracy scenarios.

The frequency with which a given partial multiplicity $l$ 
repeats in the sequence $l_{i,j}$ 
defines the partial multiplicity function $\beta_{i}(l)$. While the partial multiplicities $l_{i,j}$ partition $\alpha_i$  according to Eq.~\eqref{partial}, and hence are of an algebraic nature, the values of $\beta_{i}(l)$ partition the geometric multiplicity according to 
\begin{equation}
    \sum_{l=1}^{N}\beta_{i}(l)  = \gamma_i.
    \label{partiala}
\end{equation}
These values can hence be interpreted as geometric partial multiplicities, which count how many independent eigenvectors can be associated with the partial multiplicity $l$.
Furthermore,  Eq.~\eqref{partial} itself can be cast into the form
\begin{equation}
    \sum_{l=1}^{N}l\beta_{i}(l)  = \alpha_i.
    \label{partialb}
\end{equation}
Thereby, the partial multiplicity function $\beta_{i}(l)$ completely characterizes the algebraic and geometric structure of a given degeneracy in a convenient way.

For an EP$n$, only $\beta_{i}(n)=1$ is finite, while for an $n$-bolic point, only $\beta_{i}(1)=n$ is finite. 
For the illustrative example of an FEP with algebraic partial multiplicities (4,4,3,2,2,1),
the finite geometric partial multiplicities are $\beta_i(1)=1$, $\beta_i(2)=2$, $\beta_i(3)=1$, and $\beta_i(4)=2$, which  corresponds further to an algebraic multiplicity $\alpha_i=16$ and a geometric multiplicity $\gamma_i=6$. 

\subsection{Geometric interpretation}

In concrete geometric terms, the value $\beta_i(l)$ of the partial multiplicity function designates the
dimensionality of the eigenspace associated with a partial multiplicity $l$.
In the assignment of these spaces themselves, there exists a gauge freedom. Uniquely defined is only the space spanned by the eigenvectors of  maximal partial multiplicity \cite{Trefethen99,bid2024uniform}
\begin{equation}
l_{i,1}\equiv \ell_i. \label{eq:lmax}  
\end{equation}
The eigenvectors with maximal partial multiplicity are called the leading eigenvectors, and we will denote their geometric multiplicity $\beta_{i}(\ell_i)$ more concisely by $\beta_i$, so that  $\ell_i=l_{i,1}=l_{i,2}=\ldots=l_{i,\beta_i}$. 

For sectors with a smaller partial multiplicity, the eigenspace is defined modulo addition of eigenvectors from sectors with a larger partial multiplicity. 
This freedom can be exploited to orthogonalize the subspaces of different partial multiplicity.
The underlying hierarchical structure of these subspaces is also directly reflected in the physical signatures of FEPs, to which we turn next.

\subsection{Physical signatures}\label{sec:mathematical_cond}

The intricate features of FEPs are directly reflected in their physical signatures. 
Examining these signatures will provide the key to their  identification and classification in a systematic framework.
In particular, the response to external driving and parametric changes are both dominated by the sector of maximal partial multiplicity  $\ell_i$ \cite{Trefethen99,Graefe_2008,bid2024uniform}: 

\textit{Physical response to driving.} At an FEP,
the Green's function 
\begin{equation}
    G(E)=(E\openone-H)^{-1},
\end{equation} 
which describes the response of the system to external driving,
diverges according to 
\begin{equation}
P(E)\equiv \mathrm{tr}\,[G^\dagger(E)G(E)]\propto \frac{|\eta_i|^2}{|E-E_i|^{2\ell_i}}, 
\label{eq:superlorentzian}
\end{equation}
where $\eta_i$ is a characteristic physical response strength.

\textit{Spectral response to parameter changes.}
Upon a perturbation of strength $\varepsilon$, the degenerate eigenvalues split up into a multiplet, and become displaced by a maximal amount 
\begin{equation}
\varepsilon E_i\sim (\varepsilon\xi_i)^{1/\ell_i},
\label{eq:powerlawresponse}
\end{equation}
where $\xi_i$ is a characteristic spectral response strength.

Both response strengths are natural extensions of the behaviour at conventional EP$n$s \cite{Yoo:2011,Hashemi2022,jan1,jan2,Schomerus2024,bid2024uniform}.
The explicit expressions of these response strengths can be obtained in two approaches. With exact prior knowledge of all eigenvalues and their partial multiplicity structure, the system can be brought into a Jordan normal form, $H=TJT^{-1}$, where $T$ is a similarity transformation. Expressed in this way, the system can then be analyzed in Arnold-Jordan perturbation theory \cite{Arnold1971}. However, this perturbation theory breaks down for FEPs, while more generally the Jordan normal form construction is ill conditioned and changes  singularly under infinitesimal parameter changes. This route can therefore only be carried out in systems for which the spectral properties are exactly known. 
The second approach  \cite{bid2024uniform}, which we adopt here, starts with a systematic expansion of the Green's function around an arbitrary references energy $\Omega$, 
\begin{equation}
    G(E)=\frac{\sum_{k=0}^{N-1}(E-\Omega)^k\mathcal{B}_k}
    {\sum_{k=0}^{N}(E-\Omega)^kc_k}.
      \label{eq:mainresult0}
\end{equation}
The quantities $\mathcal{B}_k$ in the numerator are $N\times N$ matrices known as the \textit{modes of the adjugate matrix}. These modes follow directly from the Hamiltonian matrix by the  
Faddeev-LeVierre recursion relation 
\begin{align}
&\mathcal{B}_{N-1}=\openone,\quad
 &\mathcal{B}_{k-1}=A\mathcal{B}_{k}-\frac{\mathrm{tr}\,(A \mathcal{B}_{k})}{N-k}\openone,
 \label{eq:flv}
\end{align}
where $A=H-\Omega \openone$.
The modes also determine the coefficients   \begin{equation}
 c_k=-\frac{\mathrm{tr}\,(A \mathcal{B}_k)}{N-k}
 \label{eq:ckfromb}
\end{equation}
in the denominator, which are the expansion coefficients of the shifted characteristic polynomial
\begin{equation}
p(E)=\mathrm{det}\,(E\openone-H)=\mathrm{det}\,(\lambda\openone-A)
=\sum_{k=0}^{N}\lambda^k c_k\equiv q(\lambda)
,
\label{eq:shiftedq}
\end{equation}
where $\lambda=E-\Omega$.

In terms of this directly calculable data,
the response strengths are then given by \cite{bid2024uniform}
\begin{equation}
\eta_i^2=\frac{\tr(\mathcal{B}_{\alpha_i-\ell_i}^\dagger\mathcal{B}_{\alpha_i-\ell_i})}{|c_{\alpha_i}|^2}
\label{eq:etafromb}
\end{equation}
and
\begin{equation}
\xi_i^2=\frac{||\mathcal{B}_{\alpha_i-\ell_i}||_2^2}{|c_{\alpha_i}|^2},
\label{eq:xifromb}
\end{equation}
where $||\cdot||_2$ denotes the spectral norm, and the modes are calculated with $\Omega=E_i$, hence $A=H-E_i\openone$.
This algebraic approach based on the modes $\mathcal{B}_k$ therefore enables us to calculate the response strengths reliably from the Hamiltonian, without having to carry out the Jordan decomposition. 
As we will show next, the data entering the response strengths directly links to the partial multiplicity structure of FEPs (as a consequence of this connection, the two response strengths $\eta_i$ and $\xi_i$ happen to coincide for $\beta_i=1$).

\subsection{Framework for efficient analysis and design\label{sec:pepconditions}}

As we describe now,
the algebraic approach to determine the response strengths can be extended into an efficient mathematical framework that identifies and characterizes FEPs systematically and completely, and delivers explicit conditions for their formation that can be evaluated in specific models. 
To formulate this framework, we again set  $A=H-E_i\openone$, and then consider the modes $\mathcal{B}_k$ as determined by the
Faddeev-LeVierre recursion relation 
 \eqref{eq:flv}.

The main vehicle of this  characterization are two conditions. 
The eigenvalues and their algebraic multiplicity are obtained from the condition
\begin{equation}
    C_k\equiv\mathrm{tr}\,(A\mathcal{B}_k)=0, \quad k=0,1,2,\ldots,\alpha_{i}-1 ,
    \label{eq:ckcondition}
\end{equation}
originally derived in Ref. \cite{bid2024uniform} and reviewed in Appendix \ref{app:rankcondition}.
The second condition  resolves the partial multiplicities according to
\begin{equation}
\label{eq:betalresult}
\beta_i(l)=\mathrm{rnk}\,(\mathcal{B}_{\alpha_i-l-2})
-2\times \mathrm{rnk}\,(\mathcal{B}_{\alpha_i-l-1})
+\mathrm{rnk}\,(\mathcal{B}_{\alpha_i-l}),
\end{equation}
with $l\geq 1$.
We derive this additional condition, which is central to the present work, in the same Appendix \ref{app:rankcondition}.

Together, these conditions deliver the partial multiplicity function of an eigenvalue directly in terms of the shifted Hamiltonian $A$. From this information, we can reconstruct the other multiplicities as already described in subsection \ref{sec:partialmult}. Furthermore, the maximal partial multiplicity can be determined directly by the 
condition
\begin{equation}
    \mathcal{B}_k=0,\quad k=0,1,2,\ldots,\alpha_i-\ell_i-1,
    \label{eq:rankcondition2}
\end{equation}
so that the first finite mode is given 
by $\mathcal{B}_{\alpha_i-\ell_i}$, i.e., exactly by the mode that enters the algebraic expressions 
\eqref{eq:etafromb} and \eqref{eq:xifromb} for the
response strengths $\eta_i$ and $\xi_i$, respectively.
The geometric multiplicity $\beta_i$ of the leading eigenvectors is then given by the rank of this matrix,
\begin{equation}
\beta_i=\mathrm{rnk}\,(\mathcal{B}_{\alpha_i-\ell_i}),
\label{eq:beta}
\end{equation}
which results in the above-mentioned identity of the spectral strengths 
\eqref{eq:etafromb}
and 
\eqref{eq:xifromb}
if $\beta_i=1$.

All the algebraic conditions stated in this subsection  
are basis invariant, and hence can be evaluated directly from model Hamiltonians, where one can make use of the efficient recursion relation \eqref{eq:flv}.
It follows that the conditions not only serve to systematically characterize a given degenerate eigenvalue, but also can be employed to design systems that exhibit a certain desired degeneracy scenario.
With this, we are now ready to turn to specific lattice models and determine the conditions under which they exhibit FEPs in the bulk and boundary spectrum.

\section{Non-Hermitian Lieb lattices}\label{Sec:NHLieb}

Our main focus for the remainder of this work is to demonstrate the existence of  FEPs in two- and three-dimensional lattice models and to systematically incorporate these degeneracy scenarios into the body of widely studied cases of generic EPs with geometric multiplicity $\gamma_i=1$. 

\begin{figure}
    \centering
    \includegraphics[width=\columnwidth]{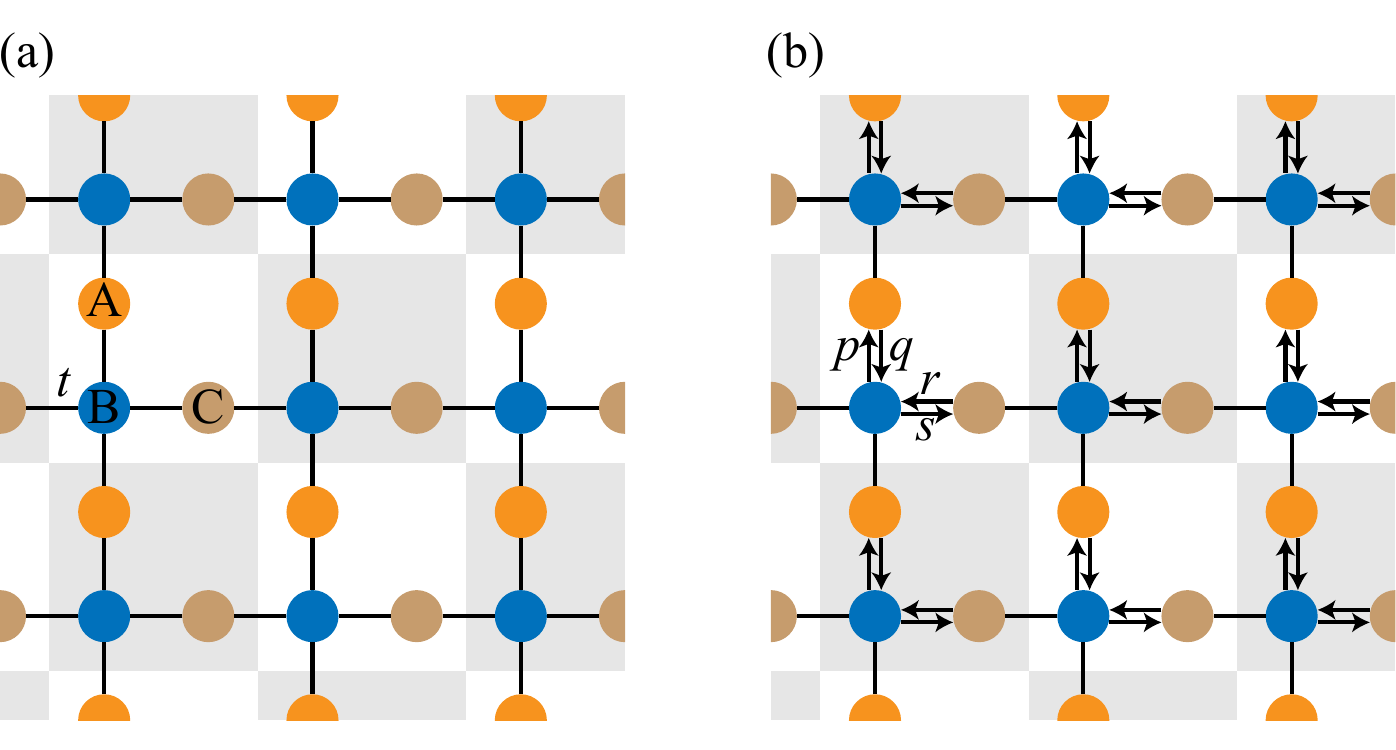}
    \caption{(a) Standard Hermitian version of the Lieb lattice, based on a three-site unit cell with uniform couplings $t$. (b) Non-Hermitian version of the model, with nonreciprocal couplings $p,q,r,s$ inside the unit cell, while the couplings between the cells remain fixed to $t$. This modification allows us to realize fragmented exceptional points in the bulk dispersion of the model.}
    \label{fig:lieblattice}
\end{figure}

We start with the Lieb lattice, 
a tight-binding model based on a unit cell of three sites, corresponding to three sublattices that we denote as A, B, and C (see Fig.~\ref{fig:lieblattice}). 
This provides the minimal setting of a system with three bands, which, hence, can display bulk degeneracies of algebraic multiplicity $\alpha=3$.
These third-order degeneracy scenarios serve as our motivation for considering this model, as $n=3$ is the first nontrivial order that can harbor an FEP, with partial multiplicities $(l_{i,1},l_{i,2})=(2,1)$. In contrast, for second-order degeneracies, there are only two scenarios: a DP within the Hermitian framework and a generic EP2 within the non-Hermitian framework.

The original Hermitian version of the Lieb lattice displays a flat band as well as two symmetric dispersive bands that meet in an $n$-bolic point of order $n=3$ (a tribolic point) \cite{lieboriginal,lieb2,lieb3}, while non-Hermitian variants have been designed to display EP$n$'s of the same order in their bulk band structure~\cite{Demange_2012,ipsita}. 
We first review these common degeneracy scenarios in the standard approach and then design a minimal model exhibiting an FEP.
Subsequently, we apply our formalism from the previous section to systematically classify these degeneracy scenarios
and identify the general conditions for Lieb lattices that support FEPs in their bulk band structure. Finally, we utilize these conditions to design a system that displays exceptional rings and lines hosting FEPs at special locations.

\label{sec:NH-Lieb}
\subsection{Hermitian model and tribolic points}
In its original Hermitian formulation, the nearest neighbors in the Lieb lattice are coupled by a uniform, real, coupling constant $t$ [see Fig.~\ref{fig:lieblattice}(a)]. Setting $t=a=1$, where $a$ is the lattice constant,
the infinite periodic system is described by the Bloch Hamiltonian 
\begin{equation}
  \mathcal{H}(\mathbf{k})=  \left(
\begin{array}{ccc}
 0 & 1+e^{i k_y} & 0 \\
 1+e^{-i k_y} & 0 & 1+e^{-i k_x} \\
 0 & 1+e^{i k_x} & 0 \\
\end{array}
\right),
\label{Hlieb}
\end{equation}
where $\mathbf{k}=(k_x,k_y)$.
The Bloch Hamiltonian reflects the lattice topology of the system, in which the A and C sites are only coupled to B sites, and vice versa. 
This constraint on the couplings induces a chiral symmetry
\begin{equation}
     \mathcal{X} \mathcal{H}(k)  \mathcal{X}=-\mathcal{H}(k),
     \label{eq:chiral}
\end{equation}
where 
\begin{equation}
  \mathcal{X}=  \left(
\begin{array}{ccc}
 1 & 0 & 0 \\
 0 & -1 & 0 \\
 0 & 0 & 1 \\
\end{array}
\right).
\end{equation}
The chiral symmetry maps any eigenstate $\mathbf{u}_i(\mathbf{k})$ with eigenvalue $E_i(\mathbf{k})$ to an eigenstate 
$ \mathcal{X}\mathbf{u}_i(\mathbf{k})$ with eigenvalue $-E_i(\mathbf{k})$. 
Therefore, the band structure is comprised of two symmetric dispersive bands $E_+(\mathbf{k})=-E_-(\mathbf{k})$, given by
\begin{equation}\label{eq:liebevals}
    E_\pm(\mathbf{k})=\pm\sqrt{2\left(\cos k_x+\cos k_y+2\right)}.
\end{equation}
By the same argument, the remaining third eigenvalue is forced to vanish throughout the Brillouin zone,
\begin{equation}
E_0(\mathbf{k})=0  \mbox{ for all $\mathbf{k}$},
\end{equation}
hence, forms a dispersionless flat band. 
A key characteristic of this flat band is its localization on the AC sublattice, while the dispersive bands exhibit
equal weights on the B and AC parts of the system.

\begin{figure}
    \centering
    \includegraphics[width=\linewidth]{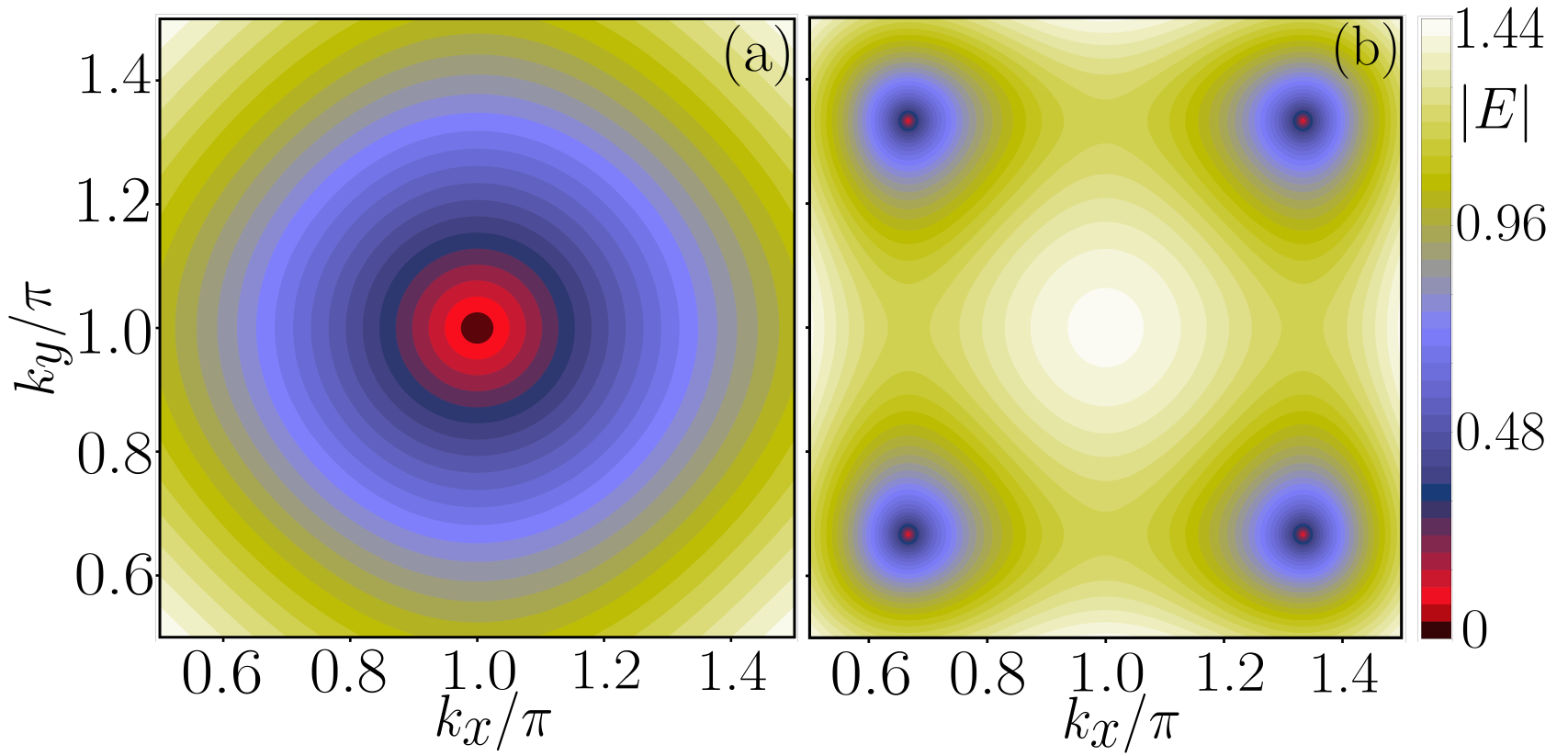}
    \caption{Contour plots of the dispersion relation $|E(\mathbf{k})|$ for (a) the conventional Hermitian Lieb lattice with Bloch Hamiltonian \eqref{Hlieb}, depicting a tribolic point at the momenta values $(k_x,k_y)=(\pi,\pi)$ around which the dispersion relation is linear, and (b) the non-Hermitian variant with Bloch Hamiltonian \eqref{nhHlieb}, proposed in Ref.~\cite{ipsita}, displaying four conventional EP$3$s with geometric multiplicity $\gamma=1$ arranged symmetrically around the point  where the tribolic point existed for the Hermitian case. The parameter $\varepsilon$ determining the strength of the non-Hermitian perturbation is set to be equal to 1. We design model systems that exhibit the more complex FEP degeneracy scenarios in their energy dispersion.}
    \label{NHlieb1}
\end{figure}

At $k_x=k_y=\pi$,
the radicand in Eq.~\eqref{eq:liebevals} vanishes, and the three bands merge at energy $E_0=E_\pm=0$, i.e., form a degeneracy with algebraic multiplicity $\alpha=3$. Furthermore, since the underlying system is Hermitian, it maintains three linearly independent eigenvectors, which then span the complete Hilbert space. Therefore, the geometric multiplicity is $\gamma=3$, and the degeneracy is a tribolic point. This scenario is illustrated in Fig.~\ref{NHlieb1}(a) in terms of a contour plot of the absolute value $|E_\pm(\mathbf{k})|$ of the dispersive bands.

\subsection{Exceptional points of order three}

As pointed out in Refs.~\cite{Demange_2012} and \cite{ipsita}, introducing a suitable non-Hermitian perturbation into the Hamiltonian
while preserving the chiral symmetry,
the tribolic point can be split into four $3^{rd}$-order exceptional points (EP3s), each characterized by an algebraic multiplicity of $\alpha=3$ and a geometric multiplicity of $\gamma=1$.
The simplest variant is given by \cite{ipsita}
\begin{equation}
  \mathcal{H}(\textbf{k}) =\left(
\begin{array}{c@{\,\,}c@{\,\,}c}
 0 &1+e^{i k_y}+i \varepsilon & 0 \\
 1+e^{-i k_y}+i \varepsilon & 0 & 1+e^{-i k_x}-i \varepsilon \\
 0 & 1+e^{i k_x}-i \varepsilon & 0 \\
\end{array}
\right)
,
\label{nhHlieb}
\end{equation}
which preserves the chiral symmetry \eqref{eq:chiral} as well as reciprocity $\mathcal{R}$, defined by the condition $\mathcal{H}(\textbf{k})=\mathcal{H}^T(-\textbf{k})$. The flat band is maintained, while the dispersive bands become
\begin{equation}
    E_\pm(\mathbf{k})=\pm \sqrt{2 \left[(1-i\varepsilon)\cos k_x+(1+i\varepsilon)\cos
   k_y-\varepsilon ^2+2\right]}.
\end{equation}
As shown in Fig.~\ref{NHlieb1}(b),  the four EP3s are symmetrically located in the Brillouin zone at momentum values $k_x=\mu k_0$, $k_y=\nu k_0$, where $k_0=\arccos\left(\frac{\varepsilon ^2}{2}-1\right)$ and $\mu,\nu=\pm1 $.
At each of these EP3s, there is only a single pair of 
right and left eigenvectors, whose nonnormalized form is given by
\begin{eqnarray}
\mathbf{u}_{\mu,\nu}=\left(
\begin{array}{c}
 i+e^{i\nu\varphi_\varepsilon} \\ 0\\
 i-e^{-i\mu\varphi_\varepsilon}
\end{array}
\right)
,\quad
\mathbf{v}_{\mu,\nu}= 
\mathbf{u}_{-\mu,-\nu}^T
,
\end{eqnarray}
where $\varphi_\varepsilon=\arccos\frac{\varepsilon}{2}$.
As a result, the geometric multiplicity is $\gamma=1$, which is a signature of an EP. Indeed, each eigenvector pair is self-orthogonal, $\mathbf{v}\mathbf{u}=0$, which is a symptom of the defectiveness of the eigensystem \cite{kato}.

\begin{figure}
    \centering
    \includegraphics[width=\linewidth]{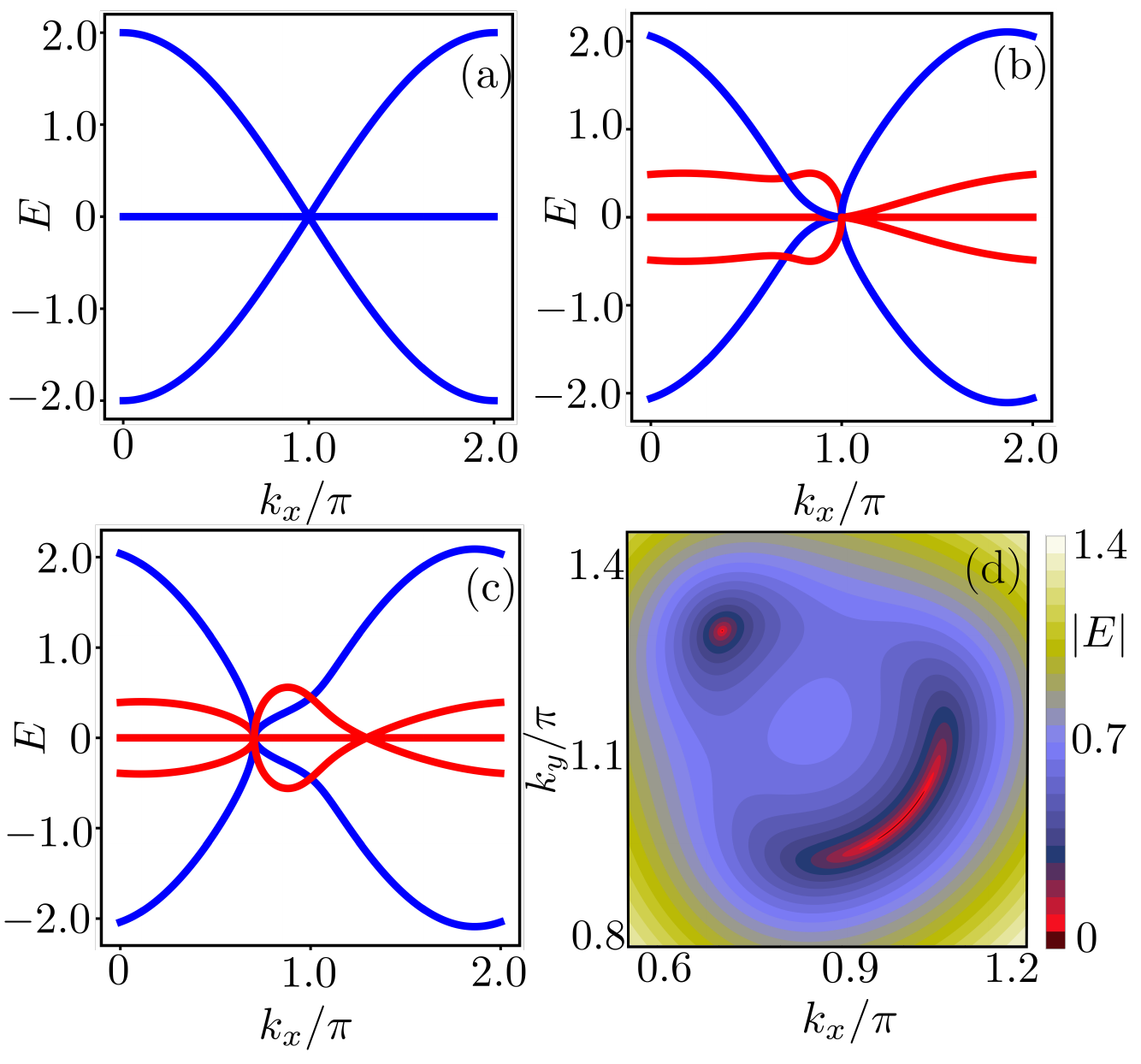}
    \caption{(a) Energy band diagram $E(k_x,k_y=\pi)$ as a function of $k_x$ for the conventional Hermitian Lieb lattice with Bloch Hamiltonian \eqref{Hlieb}, displaying the tribolic point at $k_x=\pi$. (b,c) Complex energy band diagrams
    (real part in blue, imaginary part in red)
    for the non-Hermitian Lieb lattice with Bloch Hamiltonian \eqref{nhHlieb3} and $\varepsilon=1$. In (b) $k_y=\pi$, revealing the FEP of geometric multiplicity $\gamma=2$ and partial multiplicities $(l_{1},l_{2})=(2,1)$ at $k_x=\pi$. In (c), $k_y=-2\arccot \frac{1}{2}$  is fixed to display the generic EP3 in this model, see Eq.~\eqref{momenta_values}. Panel (d) reveals the distinct  dispersion around the FEP and EP3 in this model in terms of the energy contours $|E_+(\mathbf{k})|$.  In the low energy limit, the generic EP3 displays an isotropic dispersion relation, while the dispersion relation around the FEP is highly anisotropic. 
}
    \label{NHlieb2}
\end{figure}

\subsection{Fragmented exceptional points}
To prepare our general classification in the next section, we now identify a minimal example of a non-Hermitian Lieb lattice that displays an FEP. As the system only contains three states, the FEP will have partial multiplicities $(l_1,l_2)=(2,1)$, corresponding to an algebraic multiplicity of $\alpha=3$ and a geometric multiplicity of $\gamma=2$. Therefore, the maximal partial multiplicity is $\ell=2$. In the design of the system, we are then guided by the general observation that this degeneracy structure lies in between those of a tribolic point and an EP3.
This leads us to the following choice of a minimal example,
\begin{equation}
   \mathcal{H}(\mathbf{k}) =\left(
\begin{array}{ccc}
 0 &1+e^{i k_y}+i \varepsilon & 0 \\
 1+e^{-i k_y} & 0 & 1+e^{-i k_x} \\
 0 & 1+e^{i k_x}-i \varepsilon & 0 \\
\end{array}
\right).
\label{nhHlieb3}
\end{equation}
We observe that this model preserves the chiral symmetry, but breaks reciprocity, $ \mathcal{H}^T(\mathbf{k})\neq  \mathcal{H}(-\mathbf{k})$. The flat band $E_0=0$ remains intact, while the dispersive bands are given by
\begin{equation}
    E_\pm(\mathbf{k})=\pm\sqrt{4+2\cos k_x+2\cos k_y+i\varepsilon(e^{-ik_y}-e^{-ik_x})}.
\end{equation}
Solving this relation for $E_\pm(\mathbf{k})=0$ gives us two band degeneracy points in the Brillouin zone, which are located at
\begin{equation}
k_x=k_y=\pi
\label{eq:kxkypi}
\end{equation}
and 
\begin{equation}
k_x=-k_y=2\arccot \frac{\varepsilon}{2}.
    \label{momenta_values}
\end{equation}
Therefore, unlike for the model \eqref{nhHlieb}, the tribolic point now breaks into two degeneracy points, rather than four. 
Of these, the point at $k_x=k_y=\pi$ 
 possesses two pairs of linearly independent  right  and left eigenvectors, which we may choose as
    \begin{eqnarray}
   & \mathbf{u}_1=\begin{pmatrix}
          0 \\
          0   \\
          1  
         \end{pmatrix}, \quad
         \mathbf{u}_2=\begin{pmatrix}
          1 \\
          0   \\
          0 
         \end{pmatrix}, \nonumber \\
    &\mathbf{v}_1=\begin{pmatrix}
          1,0,1
         \end{pmatrix}, \quad
         \mathbf{v}_2=\begin{pmatrix}
          0,1,0
         \end{pmatrix}.
\end{eqnarray}
This point is the desired FEP.
On the other hand, the point at $k_x=-k_y=2\arccot \frac{\varepsilon}{2}$
forms a generic EP3 with $\gamma=1$. 
Consequently, two third-order degeneracy points with different geometric multiplicities coexist in the bulk dispersion of the Hamiltonian \eqref{nhHlieb3}. 

Figure \ref{NHlieb2} illustrates the signatures of the different degeneracy scenarios in cross-sections of bulk band diagrams.
Panel (a) shows the tribolic point of the Hermitian system, while panels (b) and (c) show the FEP and EP3 in the model \eqref{nhHlieb3}, respectively. A characteristic feature of the FEP is the cusp-like shapes displayed by either the real or the imaginary parts as one approaches the degeneracy from either side. As shown in contour plots of the dispersive bands in panel (d), the dispersion around the FEP is highly anisotropic.

Based on such design-by-inspection, we can come up with further examples of Lieb lattices that display FEPs. 
However, instead of listing further such cases here, we next apply our formalism to classify the possible degeneracy scenarios systematically and completely. Afterwards, we revisit the examples given so far, and describe how they can be generalized. 

\subsection{Complete classification\label{sec:liebclassification}}

We now utilize our framework based on the modes $\mathcal{B}_k$ to
embed the examples given so far into a full classification of the possible degeneracy scenarios in the bulk dispersion relation of the Lieb lattice, and determine the general conditions for FEPs.

To carry out this classification in sufficient generality, 
we consider a  Bloch Hamiltonian of the form
\begin{equation}
  \mathcal{H}(\textbf{k}) =\left(
\begin{array}{ccc}
 0 &P & 0 \\
Q & 0 & R \\
 0 & S & 0 \\
\end{array}
\right),
\label{eq:hliebgeneral}
\end{equation}
where $P$, $Q$, $R$, and $S$ are functions of $\mathbf{k}$.
This maintains the chiral symmetry of the system, so that all degeneracies occur at energy $E=0$.
We therefore apply the modal analysis with a vanishing energy shift $\Omega=0$, so that the matrix $A=\mathcal{H}(\textbf{k})$.

\subsubsection{Modes and their rank}
Applying the Faddeev-LeVierre recursion relation  \eqref{eq:flv} we obtain
\begin{align}
    \mathcal{B}_2&=\openone,
     \nonumber 
    \\
    \mathcal{B}_1&=\mathcal{H}(\textbf{k})
&&=\left(
\begin{array}{ccc}
 0 &P & 0 \\
Q & 0 & R \\
 0 & S & 0 \\
\end{array}
\right)
,
 \nonumber   \\
    \mathcal{B}_0&=\mathcal{H}(\textbf{k})^2-\frac{\mathrm{tr}\,\mathcal{H}(\textbf{k})^2}{2}\openone
    &&=\left(
\begin{array}{ccc}
 -RS &0 & PR \\
0 & 0 & 0 \\
 QS & 0 & -PQ \\
\end{array}
\right).
\label{eq:bklieb}
\end{align}

To apply our formalism, we need to determine the rank of these matrices.
This analysis is facilitated by the fact that we can write
\begin{equation}
 \mathcal{B}_0=\left(
 \begin{array}{c}
 R \\ 0\\
 -Q
\end{array}
\right)
\left(-S,0,P\right).
\label{eq:bklieb2}  
\end{equation}

The ranks can then be read off directly from the simple form of these matrices, where we distinguish three cases:

\begin{itemize}
    \item \textbf{CASE 1}: If $(|P|+|S|)(|Q|+|R|)>0$ (hence, neither $P$ and $S$ both vanish at the same time, nor $Q$ and $R$ both vanish at the same time), then
\begin{align}
    \rnk \mathcal{B}_0=1,\quad
    \rnk \mathcal{B}_1=2,\quad
    \rnk \mathcal{B}_2=3.
    \label{eq:case1}
\end{align}
    \item \textbf{CASE 2}:
If $(|P|+|S|)(|Q|+|R|)=0$ but $|P|+|Q|+|R|+|S|>0$ (where the latter means that at least one of the quantities is finite) then
\begin{align}
    \rnk \mathcal{B}_0=0,\quad
    \rnk \mathcal{B}_1=1,\quad
    \rnk \mathcal{B}_2=3.
        \label{eq:case2}
\end{align}
    \item \textbf{CASE 3}: If
$P=Q=R=S=0$, then
\begin{align}
    \rnk \mathcal{B}_0=0,\quad
    \rnk \mathcal{B}_1=0,\quad
    \rnk \mathcal{B}_2=3.
        \label{eq:case3}
\end{align}
\end{itemize}

\subsubsection{Algebraic multiplicity}
Next, 
we determine the algebraic multiplicity of the eigenvalue from condition \eqref{eq:ckcondition}. For this, we evaluate 
\begin{align}
C_0&=\mathrm{tr}\,(\mathcal{H}(\textbf{k})\mathcal{B}_0)=0
,
\\
C_1&=\mathrm{tr}\,(\mathcal{H}(\textbf{k})\mathcal{B}_1)=2(PQ+RS),
\\
C_2&=\mathrm{tr}\,(\mathcal{H}(\textbf{k})\mathcal{B}_2)=0.
\end{align}
According to condition \eqref{eq:ckcondition}, the
algebraic degeneracy is $\alpha=1$ if $PQ+RS\neq 0$, in which case the eigenvalue is the nondegenerate eigenvalue of the flat band.
Furthermore, the 
algebraic degeneracy is $\alpha=3$
if
\begin{equation}
PQ+RS=0.
\label{eq:pqrscondition}
\end{equation}
We recover that degeneracies of order 2 are ruled out, as enforced by the symmetries of the system---but did not  need to explicitly envoke these symmetries. 
 
\subsubsection{Partial multiplicity}

With this we come to the last step of our general analysis, the determination of the partial multiplicities of the degeneracy. For this, we apply the main general result of this work, Eq.~\eqref{eq:betalresult}, to the matrices given in Eq.~\eqref{eq:bklieb}, whose ranks are specified in Eqs.~\eqref{eq:case1}, \eqref{eq:case2}, and \eqref{eq:case3}.
For completeness, we first discuss the nondegenerate scenario, and then turn to the classification and characterization of the different degenerate scenarios.

For the nondegenerate scenario, $PQ\neq -RS$, $P$ and $S$ cannot both vanish at the same time, and $R$ and $Q$ cannot both vanish at the same time. This corresponds to CASE 1, where the ranks are given by Eq.~\eqref{eq:case1}.
Since in this scenario $\alpha=1$, the multiplicity function \eqref{eq:betalresult}
reduces to the single value
\begin{equation}
\beta(1)=\mathrm{rnk}\,\mathcal{B}_0=1,
\end{equation}
while formally $\beta(2)=
\beta(3)=0$.
These values correspond to $\gamma=1$ and a single partial multiplicity $l_{1}=1$,
and thereby reproduce the
only allowed values for the multiplicity function of a nondegenerate eigenvalue.

In the degenerate scenarios, where $PQ+RS=0$ and $\alpha=3$,
the multiplicity function \eqref{eq:betalresult} follows from the
expressions
\begin{align}
\beta(1)&=\mathrm{rnk}\,\mathcal{B}_0-2\times\mathrm{rnk}\,\mathcal{B}_1+\mathrm{rnk}\,\mathcal{B}_2
,\nonumber\\
\beta(2)&=-2\times\mathrm{rnk}\,\mathcal{B}_0+\mathrm{rnk}\,\mathcal{B}_1
,\nonumber\\
\beta(3)&=\mathrm{rnk}\,\mathcal{B}_0.
\label{eq:multi3}
\end{align}

In its evaluation, we distinguish the following cases.
\begin{itemize}
\item \textbf{CASE 1}: $(|P|+|S|)(|Q|+|R|)>0$.
The rank of the modes is the same as in the nondegenerate case, i.e., given by Eq.~\eqref{eq:case1}.
Inserting these values into Eq.~\eqref{eq:multi3} we then find
\begin{align}
\beta(1)=
\beta(2)=0,\quad
\beta(3)=1.
\end{align}
These values correspond to an EP3, with a single partial multiplicity $l_{i,1}=3$ and geometric multiplicity $\gamma=1$.

\item  \textbf{CASE 2}: $(|P|+|S|)(|Q|+|R|)=0$ but $|P|+|Q|+|R|+|S|>0$.
Inserting the ranks from Eq.~\eqref{eq:case2}  into Eq.~\eqref{eq:multi3} we find
\begin{align}
\beta(1)=
\beta(2)=1,
\quad
\beta(3)=0.
\end{align}
This corresponds to an FEP with
 partial multiplicities $(l_1,l_2)=(2,1)$ and a geometric multiplicity $\gamma=2$.

\item \textbf{CASE 3}: $P=Q=R=S=0$. This is the tribolic point. The ranks are given by Eq.~\eqref{eq:case3}, from which we obtain
\begin{equation}
\beta(1)=3,\quad
\beta(2)=
\beta(3)=0.
\end{equation}
Consequently, the partial multiplicities are given by $(l_1,l_2,l_3)=(1,1,1)$.
Furthermore, the sum rule \eqref{partiala}  recovers the geometric multiplicity $\gamma=3$.    
\end{itemize}

\subsubsection{Exact general conditions for FEPs\label{sec:conditions}}

Therefore, FEPs are realized exactly when 
either (i) the quantities $P$ and $S$ both vanish and at least one of the quantities $Q$ and $R$ is finite, or (ii)
the quantities $Q$ and $R$ both vanish and at least one of the quantities $P$ and $S$ is finite. We note that the degeneracy condition $PQ+RS=0$, Eq.~\eqref{eq:pqrscondition}, is automatically fulfilled in both cases.

\subsection{Examples revisited\label{sec:examplesrevisited}}

To connect these general findings to the earlier examples, we consider a Lieb lattice with nonreciprocal couplings inside the unit cell, as shown in Fig.~\ref{fig:lieblattice}(b). This modifies the intracell couplings between the A and B sites to take values $p$ (from B to A) and $q$ (from A to B), and the intracell couplings between the B and C sites to take values $r$ (from C to B) and $s$ (from B to C). The intercell couplings maintain their reciprocal value $t$, which we again set to $t\equiv 1$.
The Bloch Hamiltonian then takes the form
of Eq.~\eqref{eq:hliebgeneral}
with 
\begin{align}
P&=p+e^{ik_y},
\nonumber
\\
Q&=q+e^{-ik_y},
\nonumber
\\
R&=r+e^{-ik_x},
\nonumber
\\
S&=s+e^{ik_x}.
\end{align}

\begin{figure*}
    \centering
    \includegraphics[width=\linewidth]{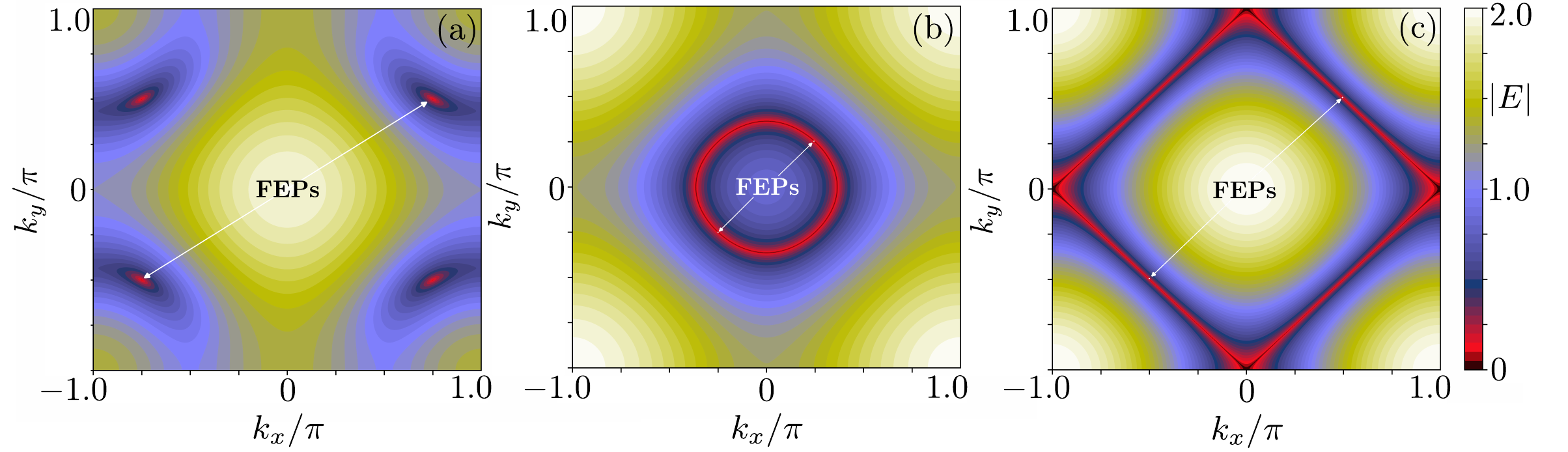}
    \caption{Contour plots of the dispersion relation $|E(\mathbf{k})|$ for the reciprocal non-Hermitian Lieb lattice model determined by the Bloch Hamiltonian \eqref{nhHliebreciprocal} for different values of the parameters $\Psi$ and $\Phi$. In panel (a), we set $\Psi=3\pi/4$ and $\Phi=\pi/2$, which realizes two FEPs at $(k_x,k_y)=(-3\pi/4,-\pi/2)$ and $(k_x,k_y)=(3\pi/4,\pi/2)$ along with two generic EP3s at $(k_x,k_y)=(-3\pi/4,\pi/2)$ and $(k_x,k_y)=(3\pi/4,-\pi/2)$. In panel (b), we set $\Phi=\Psi=\pi/4$, leading to a degenerate ring determined by the momentum relation in \eqref{eq:ring}. On this ring, two FEPs are located at $(k_x,k_y)=(\pi/4,\pi/4)$ and $(k_x,k_y)=(-\pi/4,-\pi/4)$, while the remainder is formed from generic EP3s. Setting the parameters $\Phi=\Psi=\pi/2$ in panel (c), the exceptional ring morphs into four exceptional lines, speicified in Eq.~\eqref{eq:lines}. Similar to before, the lines host two FEPs located at $(k_x,k_y)=(\pi/2,\pi/2)$ and $(k_x,k_y)=(-\pi/2,-\pi/2)$, while the remainder of the lines is formed from generic EP3s.}
    \label{fig:lieb_rings}
\end{figure*}

(1) For the original Hermitian version of the Lieb lattice, with Bloch Hamiltonian \eqref{Hlieb}, $p=q=r=s=1$. 
At the tribolic point $k_x=k_y=\pi$, $P=Q=R=S=0$, in agreement with CASE 3. We also note that these values fulfill  the degeneracy condition Eq.~\eqref{eq:pqrscondition}
trivially.

(2) For the non-Hermitian system \eqref{nhHlieb}, $p=q=1+\varepsilon$ and
$r=s=1-\varepsilon$.
The EP3s are realized at 
\begin{align}
P&=\varepsilon(e^{i\mu\varphi_\varepsilon}-i)
\nonumber
\\
Q&=\varepsilon(e^{-i\mu\varphi_\varepsilon}-i)
,
\nonumber
\\
R&=\varepsilon(e^{-i\nu\varphi_\varepsilon}+i)
\nonumber
\\
S&=\varepsilon(e^{i\nu\varphi_\varepsilon}+i),
\end{align}
where again $\varphi_\varepsilon=\arccos\frac{\varepsilon}{2}$.
These expressions fulfill Eq.~\eqref{eq:pqrscondition}
nontrivially, and as they are all finite adhere to CASE 1.

(iii) For our minimal example \eqref{nhHlieb3}, $p=1+\varepsilon$, $q=r=1$, and $s=1-\varepsilon$. At the EP3, with
$k_x=-k_y=2\arccot \frac{\varepsilon}{2}$ as specified in Eq.~\eqref{momenta_values}, the quantities $P,Q,R,S$ are again all finite and fulfill  Eq.~\eqref{eq:pqrscondition}
nontrivially, still in agreement with CASE 1. At $k_x=k_y=\pi$, $Q=R=0$ while $P$ and $S$ are finite. Accordingly, this realizes CASE 2, and hence fulfills the conditions for an FEP. 

It could now be anticipated that FEPs may arise for various situations in between the trivial and nontrival ways to fulfill the degeneracy condition $PQ+RS=0$, Eq.~\eqref{eq:pqrscondition}, of which there are four natural candidates: 
\begin{align}
&\mbox{(i)}& P=R=0,\nonumber\\
&\mbox{(ii)}& P=S=0,\nonumber\\
&\mbox{(iii)}& Q=R=0,\nonumber\\ &\mbox{(iv)}& Q=S=0, 
\label{eq:pepcandidates}
\end{align}
with the other two quantities in each case assumed to be finite (the same assumption rules out the candidates $P=Q=0$ and $R=S=0$, as the degeneracy condition then cannot be fulfilled).
However, our general analysis shows that only candidates (ii) and (iii) result in FEPs according to the conditions of CASE 2, while candidates (i) and (iv) produce additional examples of EP3s obeying the conditions of CASE 1.

\subsection{Reciprocal example with exceptional lines\label{sec:reciprocalexample}}

With these insights in mind, we conclude this section by designing a system that combines different scenarios, including for the case of exceptional lines. For this we consider  the following  model,
\begin{equation}
   \mathcal{H}(\mathbf{k}) =\left(
\begin{array}{ccc}
 0 &e^{i k_y}-e^{i\Phi} & 0 \\
 e^{-i k_y}-e^{i\Phi} & 0 & e^{-i k_x}-e^{i\Psi} \\
 0 & e^{i k_x}-e^{i\Psi} & 0 \\
\end{array}
\right),
\label{nhHliebreciprocal}
\end{equation}
with real angles $\Phi$ and $\Psi$.  Note that in contrast to the two specific non-Hermitian models \eqref{nhHlieb} and \eqref{nhHlieb3} described above, this model is reciprocal,  $\mathcal{H}(\mathbf{k})=\mathcal{H}^T(\mathbf{-k})$.
To identify the different degeneracy scenarios realized by this model, we refer to the exact general conditions for FEPs listed at the end of Sec.~\ref{sec:liebclassification}.

When the angle fulfill $\Phi\neq \Psi$ and $\Phi,\Psi \not\equiv 0 \pmod{\pi}$, the model
displays an FEP with $P=S=0$ at $(k_x,k_y)=(\Psi,\Phi)$, an FEP
with $Q=R=0$ at $(k_x,k_y)=(-\Psi,-\Phi)$, an EP3 with $Q=S=0$ at  $(k_x,k_y)=(\Psi,-\Phi)$, and an EP3 with $P=R=0$
at $(k_x,k_y)=(-\Psi,\Phi)$. This scenario is illustrated in terms of an absolute-energy band diagram in Fig. \ref{fig:lieb_rings}(a).

For the case $\Phi= \Psi \not\equiv \pi/2 \pmod{\pi}$ illustrated in Fig. \ref{fig:lieb_rings}(b), on the other hand, we find that the bulk eigenspectrum hosts an exceptional ring of algebraically three-fold degenerate eigenvalues located in the Brillouin zone.
In accordance with the degeneracy condition  $PQ+RS=0$, Eq.~\eqref{eq:pqrscondition}, this ring follows the momenta points fulfilling the relation
\begin{equation}
    \cos k_x+\cos k_y=2\cos\Phi.
    \label{eq:ring}
\end{equation}
As in the previous case, the system displays FEPs at the special positions $(k_x,k_y)=(\Phi,\Phi)$, where with $P = S = 0$, and $(k_x,k_y)=(-\Phi,-\Phi)$, where with $Q = R = 0$. All the other degenerate eigenvalues on the ring satisfying the above momentum relation are generic EP3s. 

Finally, for the case $\Phi= \Psi \equiv \pi/2 \pmod{\pi}$, Eq.~\eqref{eq:ring} reduces to
\begin{equation}
    k_y=\pm k_x.
    \label{eq:lines}
\end{equation}
The exceptional rings are hence morph into straight lines on which all the eigenvalues are algebraically three-fold degenerate, where the special locations at $(k_x,k_y)=(\pi/2,\pi/2)$, and $(k_x,k_y)=(-\pi/2,-\pi/2)$ again constitute FEPs, while all other locations on this line are generic EP3s.  
This scenario is illustrated in Fig. \ref{fig:lieb_rings}(c).

This example illustrates how our framework, based on the modes of the adjugate matrix, systematically identifies the precise and explicit conditions that result in the emergence of FEPs in the minimal case of a system with three bands. In the two following sections, we will go beyond this minimal three-band setting, and consider a three-dimensional lattice model with four bands. We will demonstrate the presence of FEPs with higher geometric multiplicity in the bulk and for the topologically protected boundary modes, where the latter manifest themselves as exceptional lines in momentum space. 

\section{Non-Hermitian higher-order Dirac semimetals}
\label{sec:NH-HODSM}
\begin{figure}[t]
    \centering
    \includegraphics[width=\columnwidth]{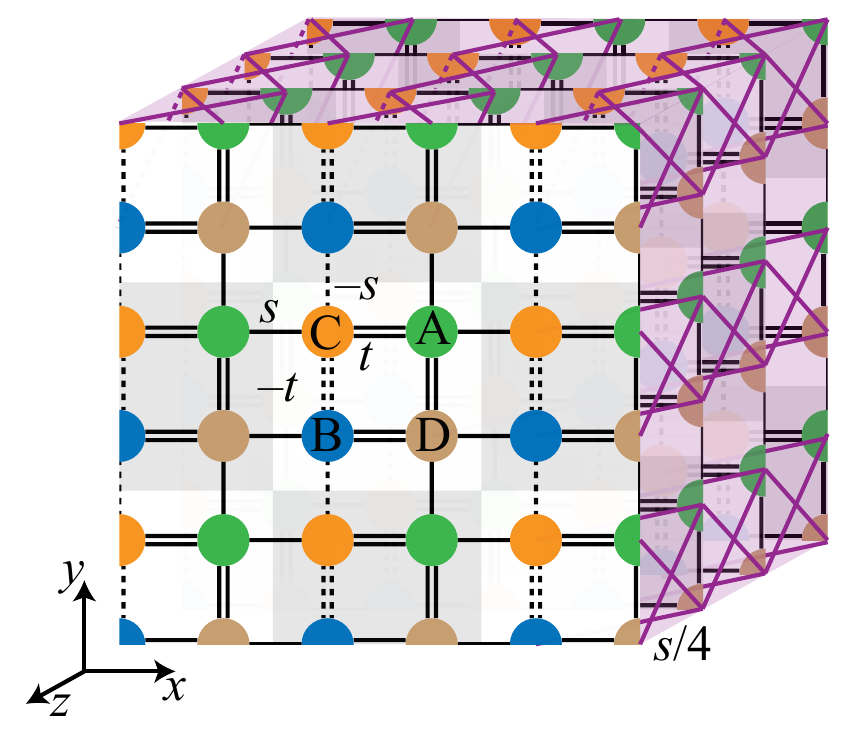}
    \caption{Lattice realization of the Hermitian higher-order Dirac semimetal, obtained from stacking two-dimensional quadrupole insulators along the $z$ direction.
    Each unit cell contains 4 sites, labeled A, B, C, and D. In each plane, the intracell and intercell couplings indicated by solid lines are $t$ and $s$,
    respectively.
The couplings between the B and C sublattices, indicated by the dashed lines, have the opposite sign, so that each plaquette carries a $\pi$ flux.
Neighboring unit cells in adjacent  planes are connected by diagonal couplings of magnitude $s/4$,
with the sign again reverted for couplings between B and C sites. 
    }
    
    \label{fig:QI}
\end{figure}

In this section, we consider the emergence of FEPs in topological four-band models based on a three-dimensional higher-order Dirac semimetal (HODSM) \cite{Hodsm}. 
Its underlying Hermitian lattice realization, depicted in Fig.~\ref{fig:QI}, is obtained by stacking two-dimensional quadrupole insulators (QIs)  \cite{multipole1} along the $z$ direction. 
As this model can host bulk degeneracies up to order $n=4$ (tetrabolic points), it provides us with more freedom to construct non-Hermitian versions displaying FEPs with various partial multiplicity structures, which is the subject of the present section. In the following section, we will further exploit that
the topologically nontrivial phase of this system, under open boundary conditions, gives rise to hinge states in its
eigenspectrum.

We first briefly discuss the model and its symmetries, and then use the modal expansion of the adjugate matrix to characterize the bulk degeneracies in the Hermitian case. 
Next, we present four non-Hermitian extensions 
of this model (see Fig. \ref{fig:HODSM_unitcell})
that host all possible degeneracy scenarios in this system, including various FEPs as well as conventional EPs and DPs (see Fig. \ref{hodsmbulkplot}). For each of these cases, we use the ranks of the modes to determine the partial multiplicity function for the encountered degeneracy scenarios.

\subsection{Hermitian parent model}\label{sec:Hodsmbulk}

\subsubsection{Construction   and symmetries}
As the foundation for creating the HODSM, we first review the lattice model of a two-dimensional QI, given by the front sheet in Fig.~\ref{fig:QI}. 
The QI consists of a square lattice with corrugated nearest-neighbor couplings, resulting in a unit cell of four sites. The intracell couplings into the $x$ and $y$  directions are of magnitude $t$, while the corresponding intercell couplings are of magnitude $s$.
In addition to this, any square plaquette (inside the unit cell, or between the unit cells) contains a $\pi$ flux, due to which one of the couplings around the plaquette is taken to be negative.  We implement this by inverting the sign of the intracell and intercell couplings connecting B and C sites, hence, along every second vertical line,  as indicated by the dashed lines in the Figure.
The Bloch Hamiltonian for this model is then given by
\begin{eqnarray}
    \mathcal{H}_{QI}(\mathbf{k})=(t+s\cos{k_x})\Gamma_4+s\sin{k_x}\Gamma_3 & 
    \nonumber \\ 
    +(t+s\cos k_y)\Gamma_2+s\sin{k_y}\Gamma_1,
    \label{HQI}
\end{eqnarray}
where the $4\times 4$ matrices  
\begin{equation}
    \Gamma_i=-\sigma_2\otimes \kappa_i \quad \text{and}  \quad \Gamma_4=\sigma_1\otimes \kappa_0
    \label{gammas}
\end{equation}
are obtained from tensor products of Pauli matrices $\sigma_i$ and $\kappa_i$, $i=1,2,3$ along with $\sigma_0=\kappa_0=\openone_2$.
The QI model described by the Hamiltonian \eqref{HQI} has spinless time-reversal symmetry $\mathcal{T}=\mathcal{K}$, where $\mathcal{K}$ denotes complex conjugation, reciprocity $\mathcal{R}$, mirror symmetries $M_x$ and $M_y$ along $x$ and $y$ direction respectively, and $C_4$ rotational symmetry $C_4\, \mathcal{H}_{QI}(k_x,k_y)\, C_4^{-1}=\mathcal{H}_{QI}(k_y,-k_x)$ with the transformation matrix
\begin{equation}
    C_4=\left(
\begin{array}{ccc}
 0 & \openone_2  \\
 -i\kappa_2 & 0 
\end{array}
\right).
\label{eq:c4}
\end{equation}
Due to the presence of $\pi$ fluxes in all plaquettes, $C_4^4=-1$. On top of the symmetries mentioned above, the Bloch Hamiltonian also possesses chiral symmetry, which again results from the division of the system into two mutually coupled sublattices. 
However, unlike for the Lieb lattice, 
the chiral operator,
\begin{equation}
   \mathcal{X}= \left(
\begin{array}{ccc}
 \openone_2 & 0  \\
 0 & -\openone_2 
\end{array}
\right)
\label{eq:chiraldm}
\end{equation}
is traceless, since the unit cell contains equal numbers of sites from these sublattices.
Therefore, the system does not exhibit a flat band.
Instead, the system possesses a symmetric band structure of two two-fold degenerate bands 
\begin{equation}
E_\pm(\mathbf{k})=\pm \sqrt{2(t^2+s^2)+2ts(\cos k_x+\cos k_y)}.
\end{equation}

\begin{figure*}
    \centering
    \includegraphics[width=\linewidth]{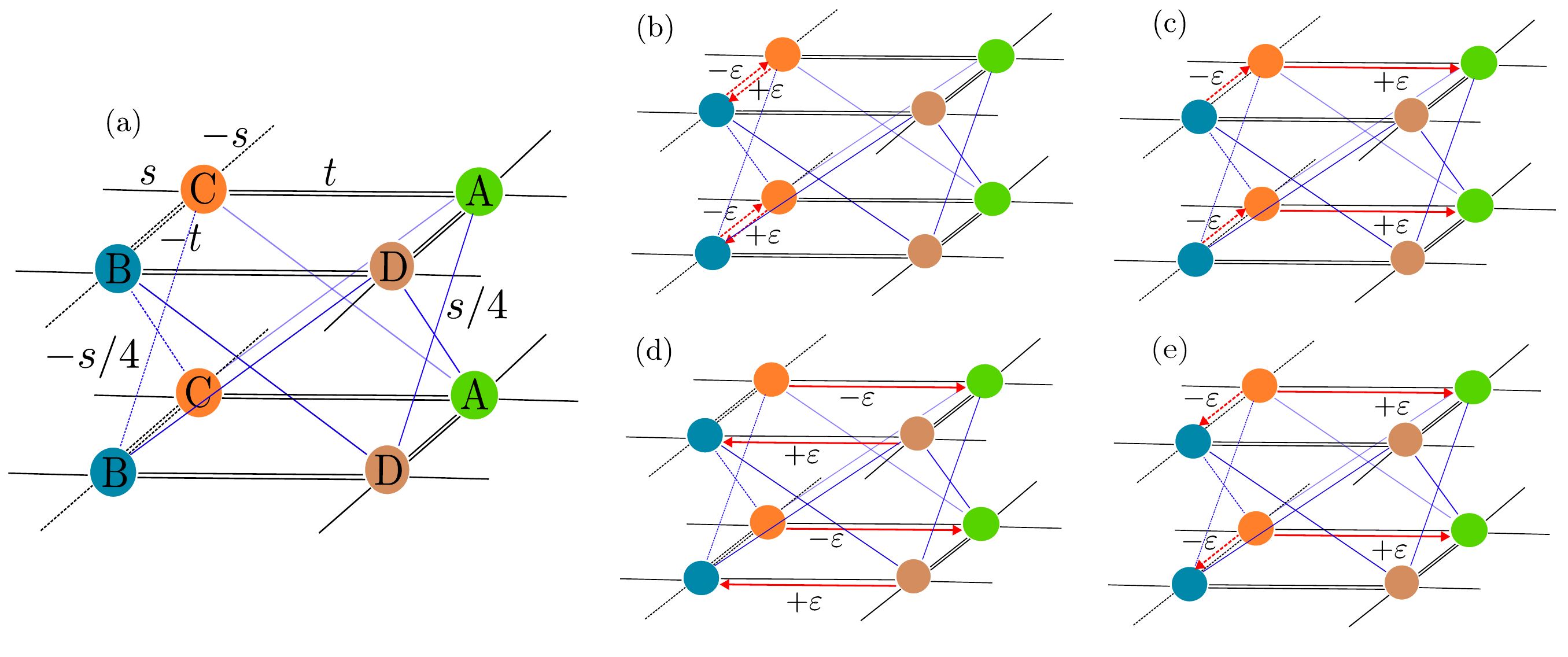}
    \caption{
    Schematics of two unit cells in the Hermitian higher-order Dirac semimetal described in Sec.~\ref{sec:Hodsmbulk}, along with its four non-Hermitian versions discussed in Secs.~\ref{subsec:model1},~\ref{subsec:model2},~\ref{subsec:model3}, and~\ref{subsec:model4}, respectively.
    Panel (a) depicts the different couplings entering the Hermitian HODSM model Hamiltonian \eqref{Hhodsm}. 
    The other panels focus on the additional non-Hermitian couplings of strength $\varepsilon$, which are all intracell and 
specified in Eq.~\eqref{eq:heps}.
The arrows denote the direction of these nonreciprocal couplings.}
    \label{fig:HODSM_unitcell}
\end{figure*}

The three-dimensional HODSM follows when one stacks the QIs and interconnects them with intracell criss-cross couplings [see again Fig.~\ref{fig:QI}, as well as Fig.~\ref{fig:HODSM_unitcell}(a)], set here for convenience to $s/4$.
In the Bloch Hamiltonian \eqref{HQI} this renormalizes the intracell coupling parameter  $t\rightarrow t+\frac{s}{2}\cos{k_z}$. 
The HODSM Bloch Hamiltonian is therefore given by
\begin{equation}
    \mathcal{H}(\mathbf{k})=  \sum_{j=1}^4 d_j \Gamma_j,
    \label{Hhodsm}
\end{equation}
with $d_{1/3}=s \sin k_{y/x}$, and $ d_{2/4}=(t +\frac{s}{2}\cos k_z+ s\cos k_{y/x})$. This model preserves all the symmetries of the QI model mentioned earlier. 
We next discuss the bulk energy bands and characterize the degeneracy scenarios occurring in the Brillouin zone.

\subsubsection{Modal analysis of bulk degeneracies}

To prepare the application of our formalism to non-Hermitian extensions of this model, we now apply the modal expansion to the Hermitian case.
For this, we note that the Bloch Hamiltonian   \eqref{Hhodsm} can be  written in the block off-diagonal form
\begin{equation}
    \mathcal{H}(\mathbf{k})=\left(
\begin{array}{ccc}
 0 & P(\mathbf{k})  \\
 P^{\dagger}(\mathbf{k}) & 0 
\end{array}
\right),
\label{Hodsm2}
\end{equation}
where the blocks are  $2\times2$ matrices
\begin{equation}
P(\mathbf{k})=p_0(\mathbf{k})\sigma_0+p_1(\mathbf{k})\sigma_1+p_2(\mathbf{k})\sigma_2+p_3(\mathbf{k})\sigma_3
    \label{eq:A_form}
\end{equation}
with
\begin{align}
p_0(\mathbf{k})&=t+s\cos{k_x}+\frac{s}{2}\cos k_z,\nonumber
\\
p_1(\mathbf{k})&=i\,s\sin{k_y},\nonumber
\\
p_2(\mathbf{k})&=i\left(t+s\cos{k_y}+\frac{s}{2}\cos k_z\right),\nonumber
\\
p_3(\mathbf{k})&=i\,s\sin{k_x}.
\label{eq:coefficient_a}
\end{align}
The bulk dispersion relation then takes the form
\begin{equation}
    E_\pm(\mathbf{k})=\pm \sqrt{\sum_{i=0}^{3}|p_i(\mathbf{k})|^2},
    \label{eq:HODSM_dispersion}
\end{equation}
still describing two doubly-degenerate bands.
The form of this expression implies that four-fold degenerate band-touching points appear when all quantities $p_i(\mathbf{k})$ vanish simultaneously, i.e., when
\begin{equation}    p_0(\mathbf{k})=p_1(\mathbf{k})=p_2(\mathbf{k})=p_3(\mathbf{k})=0.
\label{eq:pallvanish}
\end{equation}
Within the first Brillouin zone, the location of these algebraically four-fold degenerate points therefore occur at the momentum values \begin{equation}
    (k_x,k_y,k_z)=(0,0,\pm\arccos\left(-2-\frac{2t}{s}\right))
    \label{eq:hermitiandiracpoints1}
\end{equation} 
as well as 
\begin{equation}
   (k_x,k_y,k_z)=(\pi,\pi,\pm\arccos\left(2-\frac{2t}{s}\right)),
    \label{eq:hermitiandiracpoints2}
\end{equation} 
where the two cases effectively can be mapped onto each other by inverting the sign of $t/s$. In Fig.~\ref{hodsmbulkplot}(a) we illustrate these degeneracies by plotting the dispersion relation  \eqref{eq:HODSM_dispersion} as a function of $k_z$ for $k_x=k_y=0$  and $s=-t=1$, so that the bands touch at $k_z=\pm \pi/2$.

To analyze these four-fold bulk degeneracies at $E=0$, we will therefore fix $k_x=k_y=0$, and in the modal formalism we set the energy shift parameter $\Omega=0$, so that the matrix $A=\mathcal{H}(\mathbf{k})$. This analysis is further facilitated by the identities  
\begin{align}
    \mathcal{H}^2(\mathbf{k})&=E^2(\mathbf{k}) \,\openone,\nonumber
\\
\mathrm{tr}\,\mathcal{H}^2(\mathbf{k})&=4E^2(\mathbf{k}),
\label{eq:trace_cond}
\end{align}
where we denote the common value of the squared dispersion $E_+^2(\mathbf{k})=E_-^2(\mathbf{k})\equiv E^2(\mathbf{k})$ (these relations are tied to the alternative interpretation of the system as a square-root topological insulator \cite{Arkinstall2017}, which we revisit in the discussion of hinge states).
The modes calculated using the Faddeev-LeVierre recursion relation  \eqref{eq:flv} are then given by
\begin{align}
\label{eq:bkHODSM}
    \mathcal{B}_3&=\openone,
    \\
    \mathcal{B}_2&=\mathcal{H}(\textbf{k}) 
,
 \nonumber   \\
    \mathcal{B}_1&=\mathcal{H}^2(\textbf{k})-\frac{\mathrm{tr}\,\mathcal{H}^2(\textbf{k})}{2}\openone
    &&= -\frac{\mathrm{tr}\,\mathcal{H}^2(\mathbf{k})}{4}\openone,\nonumber\\
    \mathcal{B}_0&=\mathcal{H}(\textbf{k}) \mathcal{B}_1-\frac{\mathrm{tr}\,\mathcal{H}(\textbf{k})\mathcal{B}_1}{3}\openone
    &&= -\mathcal{H}(\mathbf{k})\frac{\mathrm{tr}\,\mathcal{H}^2(\mathbf{k})}{4}\openone.
\nonumber
\end{align}
Given the condition \eqref{eq:pallvanish}
and the form \eqref{Hodsm2} of the Bloch Hamiltonian, $A=\mathcal{H}(\mathbf{k})$ vanishes at the bulk degeneracy point.

We first use Eq.~\eqref{eq:ckcondition} to verify consistency with the algebraic multiplicity of the degenerate eigenvalue.
As $A=0$, all coefficients $C_i=\mathrm{tr} \,(A\mathcal{B}_i)$, $i=0,1,2,3$, appearing in the condition vanish. This implies that the algebraic multiplicity takes the value $\alpha_i=4$, which recovers the actual algebraic multiplicity of this degeneracy.

Next, we resolve the partial multiplicities by using Eq.~\eqref{eq:betalresult}. For the present model with dimension $N=4$, this gives the relations
\begin{align}
\beta(1)&=\mathrm{rnk}\,\mathcal{B}_1-2\times\mathrm{rnk}\,\mathcal{B}_2+\mathrm{rnk}\,\mathcal{B}_3
,\nonumber\\
\beta(2)&=\mathrm{rnk}\,\mathcal{B}_0-2\times\mathrm{rnk}\,\mathcal{B}_1+\mathrm{rnk}\,\mathcal{B}_2
,\nonumber\\
\beta(3)&=-2\times\mathrm{rnk}\,\mathcal{B}_0+\mathrm{rnk}\,\mathcal{B}_1
,\nonumber\\
\beta(4)&=\mathrm{rnk}\,\mathcal{B}_0.
\label{eq:multitetrabolic}
\end{align}

Applied to the present degeneracy, where the Hamiltonian itself vanishes, the only mode with nonvanishing rank is $\mathcal{B}_3$. Therefore, we have
\begin{equation}
    \beta(1)=4,\quad\beta(2)=
\beta(3)=
\beta(4)=0.
\end{equation}
As required, this describes a tetrabolic point, in which each eigenvalue has partial multiplicity 1.

Indeed, according to condition \eqref{eq:rankcondition2}, since the first finite mode carries the index $\alpha_i-\ell_i=3$, the maximal partial multiplicity is $\ell_i=1$, as required for all degeneracies of Hermitian systems. The geometric multiplicity of the corresponding leading eigenvectors then also follows from Eq.~\eqref{eq:beta}
as $\beta_i=\mathrm{rnk}\,(\mathcal{B}_3)=4$. 
Finally, we also recover the total geometric multiplicity from Eq.~\eqref{partiala} as 
    \begin{equation}
    \gamma_i=\sum_{l=1}^{N}\beta_{i}(l)  =4.
\end{equation}

This concludes the analysis of the tetrabolic points in the bulk energy eigenspectrum of the Hermitian HODSM model.
In the following subsections, we construct various non-Hermitian versions of this model and demonstrate the existence of FEPs in their bulk energy dispersion relations.

\subsection{Overview of non-Hermitian model variants }\label{sec:Hodsmnhbulk}

Compared with the example of the Lieb lattice described in Sec. \ref{Sec:NHLieb}, where we could only construct FEPs with partial multiplicities $(l_{i,1},l_{i,2})=(2,1)$,
the maximal algebraic multiplicity $\alpha_i=4$ displayed by degeneracies in the  HODSM model gives us more freedom to construct FEPs with distinct partial multiplicities beyond the generic EP.
Therefore, we will now describe non-Hermitian variants of this system leading to FEPs with the three possible sets of partial multiplicities $(l_{i,1},l_{i,2})=(3,1) , \,(2,2)$ and $(l_{i,1},l_{i,2},l_{i,3})=(2,1,1)$. 
Furthermore, we will encounter generic EP4s with $(l_{i,1})=(4)$, as well as two-fold degenerate diabolic points with $(l_{i,1},l_{i,2})=(1,1)$.
This covers all possible degeneracy scenarios that are compatible with the chiral symmetry of this model.

We base these variations on the non-Hermitian extension
\begin{equation}
    \mathcal{H}(\mathbf{k})=\left(
\begin{array}{ccc}
 0 & Q(\mathbf{k})  \\
 R(\mathbf{k}) & 0 
\end{array}
\right),
\label{eq:HODSMNH}
\end{equation}
of the Bloch Hamiltonian \eqref{Hodsm2},
where the matrices
\begin{align}Q(\mathbf{k})&=q_0(\mathbf{k})\sigma_0+q_1(\mathbf{k})\sigma_1+q_2(\mathbf{k})\sigma_2+q_3(\mathbf{k})\sigma_3,
\label{eq:qexpansion}
\\
R(\mathbf{k})&=r_0(\mathbf{k})\sigma_0+r_1(\mathbf{k})\sigma_1+r_2(\mathbf{k})\sigma_2+r_3(\mathbf{k})\sigma_3
\label{eq:rexpansion}
\end{align}
are now no longer constrained by the condition $R^\dagger(\mathbf{k})=Q(\mathbf{k})$, hence, in general are independent of each other.

The non-Hermitian extension \eqref{eq:HODSMNH}
retains the block structure and resulting chiral symmetry of the Bloch Hamiltonian, 
leading to a generally complex dispersion relation of the form 
\begin{align}
E_{\pm,\pm} &=\pm\frac{1}{\sqrt{2}}\left(\tr [RQ]
\pm\sqrt{\tr^2 [RQ]-4\det R\det Q}
\right)^{1/2}
\nonumber
\\
&=\pm\left(\sum_{i=0}^3 r_i q_i
\pm\sqrt{\sum_{i<j}^3 g_i(g_iq_ir_j +r_iq_j)^2 } \right)^{1/2}
,
\end{align}
where the two signs can be chosen independently, 
$g_{0,1,2,3}=(1,-1,-1,-1)$,
and we dropped the momentum dependence for conciseness. 
We will utilize this block structure to obtain degeneracies at $E=0$. We read off that such degeneracies occur at momentum values for which
\begin{equation}
\det Q(\mathbf{k}) = 0 \quad \mbox{or}\quad \det  R(\mathbf{k}) = 0.
\label{eq:nonhermitianquant}
\end{equation}
Furthermore, these degeneracies have algebraic multiplicity 2 if $\tr [Q(\mathbf{k}) R(\mathbf{k})] \neq 0$, and algebraic multiplicity 4 if $\tr [Q(\mathbf{k}) R(\mathbf{k})] = 0$. 

Specifically, we will consider minimal extensions that can be obtained by modifying the existing couplings in the parent model.
An overview of the considered coupling modifications is given in Fig.~\ref{fig:HODSM_unitcell}, where the non-Hermiticity parameter is again denoted by $\varepsilon$.
This leads to four models of the specific form
\begin{equation}
    \mathcal{H}^{(n)}(\mathbf{k})=\left(
\begin{array}{ccc}
 0 & P(\mathbf{k})  \\
 P^\dagger(\mathbf{k}) & 0 
\end{array}
\right)
+h^{(n)},
\label{eq:HODSMNHspecific}
\end{equation}
where 
\begin{align}
h^{(1)}&=\left(
\begin{array}{cccc}
 0 & 0 & 0 & 0 \\
 0 & 0 & \varepsilon  & 0 \\
 0 & -\varepsilon  & 0 & 0 \\
 0 & 0 & 0 & 0 \\
\end{array}
\right), &
h^{(2)}&=\left(
\begin{array}{cccc}
 0 & 0 & \varepsilon & 0 \\
 0 & 0 & 0  & 0 \\
 0 & -\varepsilon  & 0 & 0 \\
 0 & 0 & 0 & 0 \\
\end{array}
\right),
\nonumber
\\
h^{(3)}&=\left(
\begin{array}{cccc}
 0 & 0 & -\varepsilon & 0 \\
 0 & 0 & 0  & \varepsilon \\
 0 & 0  & 0 & 0 \\
 0 & 0 & 0 & 0 \\
\end{array}
\right),
&
h^{(4)}&=
\left(
\begin{array}{cccc}
 0 & 0 & \varepsilon & 0 \\
 0 & 0 & -\varepsilon  & 0 \\
 0 & 0  & 0 & 0 \\
 0 & 0 & 0 & 0 \\
\end{array}
\right).
\label{eq:heps}
\end{align}
Where convenient, we refer to these four models as NH model 1, 2, 3, and 4.
As in the Hermitian case, we
designed these modifications so that the degeneracies occur at $k_x=k_y=0$. 
This allows us to focus on the role of the momentum component $k_z$ and the non-Hermiticity parameter $\varepsilon$. 
We now discuss the different scenarios obtained in this way in detail, where we will set $s=-t\equiv 1$ to simplify expressions. 

\subsection{Non-Hermitian model 1:\\ Generic exceptional points and diabolic points}\label{subsec:model1}
We start with a scenario that yields generic exceptional points of order 4 (EP4s), along with diabolic points of algebraic and geometric degeneracy 2. We achieve this by setting the expansion coefficients in Eqs.~\eqref{eq:qexpansion} and \eqref{eq:rexpansion} to
\begin{align}
    &q_0^{(1)}(\mathbf{k})= p_0(\mathbf{k})\nonumber \\
    & q_1^{(1)}(\mathbf{k})= p_1(\mathbf{k}) +\varepsilon/2, \nonumber \\
    & q_2^{(1)}(\mathbf{k})= p_2(\mathbf{k}) -i\varepsilon/2, \nonumber \\
    & q_3^{(1)}(\mathbf{k})= p_3(\mathbf{k}),
    \label{eq:upper_a}
\end{align}
as well as 
\begin{align}
    & r_0^{(1)}(\mathbf{k})=p_0^*(\mathbf{k}), \nonumber \\
    & r_1^{(1)}(\mathbf{k})= p_1^*(\mathbf{k}) -\varepsilon/2,  \nonumber \\
    & r_2^{(1)}(\mathbf{k})= p_2^*(\mathbf{k}) -i\varepsilon/2,\nonumber \\
    &r_3^{(1)}(\mathbf{k})=p_3^*(\mathbf{k}),
    \label{eq:lower_b}
\end{align}
where $p_i(\mathbf{k})$ are the expansion coefficients of the Hermitian parent model specified in Eq.~\eqref{eq:coefficient_a}, and the corresponding coupling modifications are depicted in Fig.~\ref{fig:HODSM_unitcell}(b).
Translated into Eq.~\eqref{eq:HODSMNHspecific}, these choices realize NH model 1. Diagonalizing the resulting Hamiltonian $\mathcal{H}^{(1)}(\mathbf{k})$ gives the dispersion relation
\begin{equation}
    E=\pm\frac{1}{\sqrt{2}}\left(\cos^2 k_z-\varepsilon^2\pm\varepsilon  \sqrt{\varepsilon^2-\cos^2 k_z}\right)^{1/2},
    \label{eq:dis1}
\end{equation}
where we have set $k_x=k_y=0$ and $s=-t\equiv 1$ as discussed previously. 
Upon setting the above equation to zero, we obtain degenerate eigenvalues in the first Brillouin zone  at
\begin{equation}
    k_z=\pm \frac{\pi}{2},\quad k_z=\pm \frac{1}{2} \arccos\left(2 \varepsilon ^2-1\right).
    \label{disgm1}
\end{equation}
We use our modal analysis to resolve the partial multiplicities of the above degenerate scenarios, and start by setting $k_z=\pm \frac{1}{2} \arccos\left(2 \varepsilon ^2-1\right)$, which corresponds to four locations in the first Brillouin zone. At these values, the Bloch Hamiltonian takes the explicit form
\begin{equation}
   \mathcal{H}= \left(
\begin{array}{cccc}
 0 & 0 & \varepsilon/2 &  \varepsilon/2 \\
 0 & 0 &  \varepsilon/2 &  \varepsilon/2 \\
 \varepsilon/2 & 3\varepsilon/2 & 0 & 0 \\
 \varepsilon/2 & \varepsilon/2 & 0 & 0 \\
\end{array}
\right).
\end{equation}
Applying the Faddeev-LeVierre recursion relation  \eqref{eq:flv} with $\Omega=0$, $A=\mathcal{H}$ gives us  the  modes
\begin{align}
    \mathcal{B}_3&=\openone,
     \nonumber 
    \\
    \mathcal{B}_2&=\mathcal{H}
,
 \nonumber   \\
    \mathcal{B}_1&
    =
 \frac{\varepsilon^2}{2}   \left(
 \begin{array}{cccc}
1 & -1 & 0 & 0 \\
 1 & -1 & 0 & 0 \\
 0 & 0 & -1 & -1 \\
 0 & 0 & 1 & 1 \\
\end{array}
\right),\nonumber\\
    \mathcal{B}_0
    &=
     \frac{\varepsilon^3}{2}  
    \left(
\begin{array}{cccc}
 0 & 0 & 0 & 0 \\
 0 & 0 & 0 & 0 \\
 -1 & 1 & 0 & 0 \\
 1 & -1 & 0 & 0 \\
\end{array}
\right),
\label{eq:bkHODSMNH1}
\end{align}
along with their ranks
\begin{align}
    \rnk \mathcal{B}_0=1,\quad
    \rnk \mathcal{B}_1=2,\quad
    \rnk \mathcal{B}_2=3,\quad
    \rnk \mathcal{B}_3=4.
\end{align}
These then deliver the partial multiplicity functions of the considered degeneracy scenario from relation \eqref{eq:multitetrabolic} as
\begin{equation}
    \beta(4)=1,\quad \beta(1)=
\beta(2)=
\beta(3)=0.
\end{equation}
Therefore, our formalism demonstrates that we have indeed encountered a generic EP4.

At $k_z=\pm \pi/2$, 
the Bloch Hamiltonian reduces to
\begin{equation}
    \mathcal{H}=\left(
\begin{array}{cccc}
 0 & 0 & 0 & 0 \\
 0 & 0 & \varepsilon  & 0 \\
 0 & -\varepsilon  & 0 & 0 \\
 0 & 0 & 0 & 0 \\
\end{array}
\right)\equiv h^{(1)}.
\label{eq;h1eps}
\end{equation}
Repeating the modal analysis for this case, 
we obtain
\begin{align}
    \mathcal{B}_3&=\openone,
     \nonumber 
    \\
    \mathcal{B}_2&=\mathcal{H}
,
 \nonumber   \\
    \mathcal{B}_1&
    =
    \left(
\begin{array}{cccc}
 \varepsilon^2 & 0 & 0 & 0 \\
 0 & 0 & 0 & 0 \\
 0 & 0 & 0 & 0 \\
 0 & 0 & 0 & \varepsilon ^2 \\
\end{array}
\right),\nonumber\\
    \mathcal{B}_0
&=0.
\end{align}

From this, we compute their ranks
\begin{align}
    \rnk \mathcal{B}_0=0,\quad
    \rnk \mathcal{B}_1=\rnk \mathcal{B}_2=2,\,\,\text{and}\,\,
    \rnk \mathcal{B}_3=4.
\end{align}
Noticing that $C_2=\mathrm{tr} \,(\mathcal{H}\mathcal{B}_2)$ is now finite, while $C_1= \mathrm{tr} \,(\mathcal{H}\mathcal{B}_1)=0$ still vanishes, condition \eqref{eq:ckcondition}  delivers an algebraic degeneracy $\alpha_i=2$. To determine whether this is an EP or DP, we again obtain the partial multiplicity function from the relations \eqref{eq:betalresult}, which now take the  form
\begin{align}
\beta(1)&=-2\times\mathrm{rnk}\,\mathcal{B}_0+\mathrm{rnk}\,\mathcal{B}_1=2,\nonumber\\
\beta(2)&=\mathrm{rnk}\,\mathcal{B}_0=0.
\label{eq:secondorderbetas}
\end{align}
Therefore, remarkably, the degeneracy scenario at $k_z=\pm \pi/2$ is a DP, with two independent eigenvectors of partial multiplicity 1.

\begin{figure*}
    \centering
    \includegraphics[width=\linewidth]{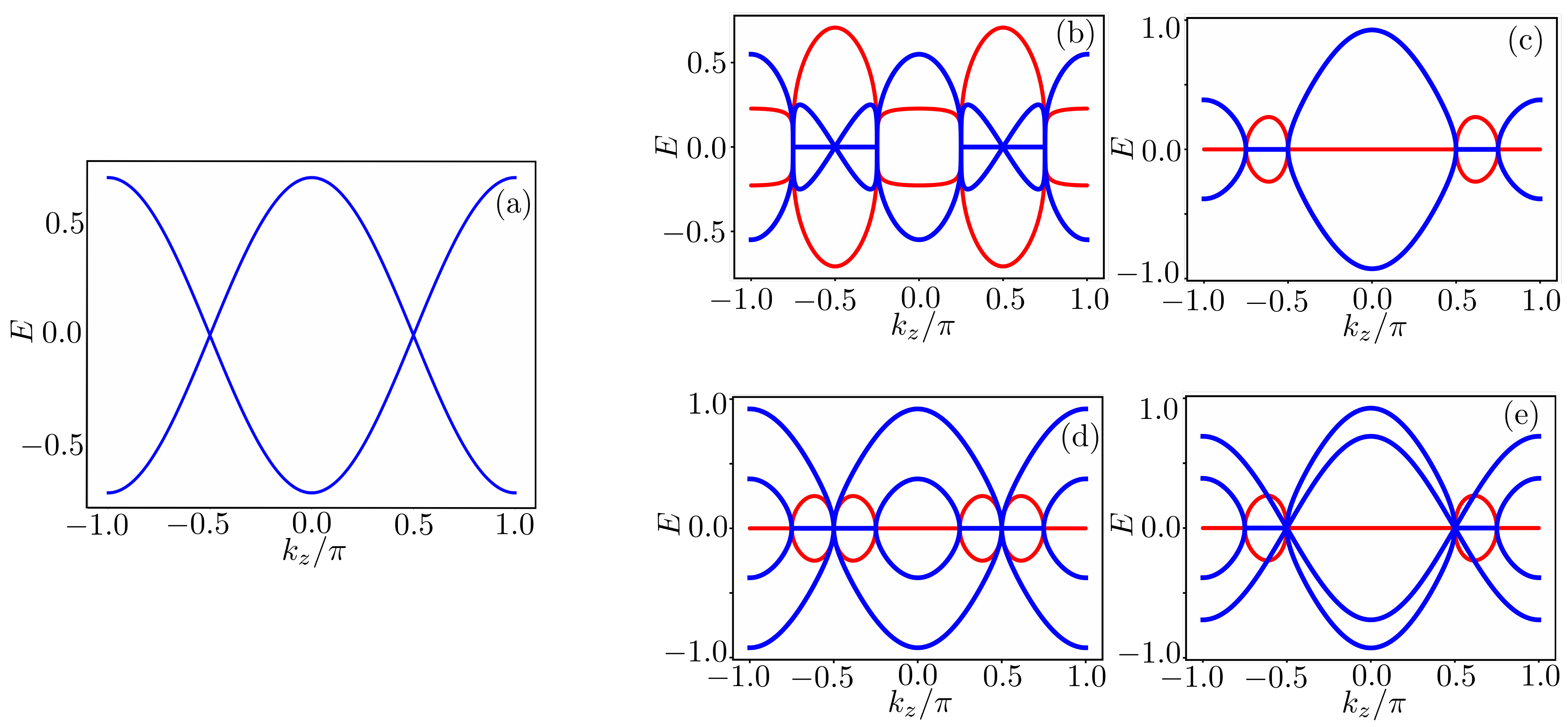}
    \caption{Bulk energy bands $E(\mathbf{k})$ as a function of $k_z$ for $k_x=k_y=0$ for  (a) the Hermitian HODSM specified by the Bloch Hamiltonian 
\eqref{Hhodsm}, as well as (b-e) the non-Hermitian variants with  additional non-Hermitian couplings of strength $\varepsilon$ 
specified in Eq.~\eqref{eq:heps} (real parts shown in blue, imaginary parts shown in red). In all panels, the couplings from the Hermitian parent model are set to $s=-t=1$. The energy bands in panel (a) follow from Eq.~\eqref{eq:HODSM_dispersion} and depict two four-fold degenerate tetrabolic points at $k_z=\pm \pi/2$.
    Panel (b) shows the dispersion relation  \eqref{eq:dis1} of NH model 1 with  $\varepsilon=1/\sqrt{2}$, displaying two DPs at $k_z=\pm \pi/2$ as well as  four generic EP$4$s at $k_z=\pm \pi/4$ and $k_z=\pm 3\pi/4$.  Panel (c) shows
    the dispersion relation \eqref{eq:dis2} of NH model 2 with $\varepsilon=1/\sqrt{2}$, 
    displaying four FEPs with partial multiplicities $(l_{i,1},l_{i,2})=(3,1)$ at $k_z=\pm\pi/2$ and $k_z=\pm3\pi/4$.
Panel (d) shows
    the dispersion relation \eqref{eq:dis3} of NH model 3 with $\varepsilon=1/2$, resulting in two
    FEPs with   $(l_{i,1},l_{i,2})=(2,2)$ located at $k_z=\pm \pi/2$, as well as four generic EP$2$s at $k_z=\pm \pi/4,$ and $k_z= \pm 3\pi/4$.
Lastly,  panel (e) shows the dispersion relation \eqref{eq:dis4} of NH model 4  for $\varepsilon=1/\sqrt{8}$, displaying two FEPs with $(l_{i,1},l_{i,2},l_{i,3})=(2,1,1)$ at $k_z=\pm \pi/2$ and two  generic EP$2$s at $k_z=\pm 3\pi/4$. }
    \label{hodsmbulkplot}
\end{figure*}

We plot the dispersion relation \eqref{eq:dis1} in Fig. \ref{hodsmbulkplot}(b) for $\varepsilon=1/\sqrt{2}$, illustrating the coexistence of the pair of diabolic points and the four EP4s in the bulk energy bands of this model.

\subsection{Non-Hermitian model 2: FEPs with\\ geometric multiplicity $\gamma_i=2$ and $(l_{i,1},l_{i,2})=(3,1)$}\label{subsec:model2}
For degeneracy scenarios with algebraic multiplicity $\alpha_i=4$ and geometric multiplicity $\gamma_i=2$, we have two possible FEPs with partial multiplicities $(l_{i,1},l_{i,2})=(3,1)$ and $(l_{i,1},l_{i,2})=(2,2)$. We realize these in two separate models, covered in the present and the following subsection, where we here deal with the case $(l_{i,1},l_{i,2})=(3,1)$.

Starting again from the general form \eqref{eq:HODSMNH} of the non-Hermitian Hamiltonian, 
we set the expansion coefficients to
\begin{align}
    &q_0^{(2)}(\mathbf{k})= p_0(\mathbf{k})+\varepsilon/2\nonumber \\
    & q_1^{(2)}(\mathbf{k})= p_1(\mathbf{k}), \nonumber \\
    & q_2^{(2)}(\mathbf{k})= p_2(\mathbf{k}), \nonumber \\
    & q_3^{(2)}(\mathbf{k})= p_3(\mathbf{k})+\varepsilon/2,
    \label{eq:upper_a2}
\end{align}
while keeping the coefficients $r_i^{(2)}(\mathbf{k})$ of the lower block the same as in the previous model, see relations \eqref{eq:lower_b}.
In terms of Eq.~\eqref{eq:HODSMNHspecific}, these choices realize NH model 2.
The dispersion relation is then given as
\begin{equation}
    E_\pm=\pm \frac{1}{\sqrt{2}} \sqrt{\cos(k_z)[\varepsilon+\cos(k_z)]},
    \label{eq:dis2}
\end{equation}
where each band is two-fold degenerate. Four-fold degeneracies now occur at momentum values
\begin{equation}
    k_z=\pm\pi/2\quad \text{and} \quad k_z=\pm\arccos(-\varepsilon).
\end{equation}
We begin our modal analysis with the degeneracy at $k_z=\pm\pi/2$,
where the Hamiltonian takes the simple form
\begin{equation}
    \mathcal{H}=\left(
\begin{array}{cccc}
 0 & 0 & \varepsilon & 0 \\
 0 & 0 & 0  & 0 \\
 0 & -\varepsilon  & 0 & 0 \\
 0 & 0 & 0 & 0 \\
\end{array}
\right)\equiv h^{(2)}.
\label{eq:HODSMNH2a}
\end{equation}
The modes follow as
\begin{align}
    \mathcal{B}_3&=\openone,
     \nonumber 
    \\
    \mathcal{B}_2&=\mathcal{H},
\\
    \mathcal{B}_1&=\left(
\begin{array}{cccc}
 0 & -\varepsilon^2 & 0 & 0 \\
 0 & 0 & 0  & 0 \\
 0 & 0  & 0 & 0 \\
 0 & 0 & 0 & 0 \\
\end{array}
\right),
\nonumber\\
    \mathcal{B}_0&=0,
\label{eq:bkHODSMNH2a}
\end{align}
and substituting their ranks
\begin{equation}
    \rnk \mathcal{B}_0=0,\quad
    \rnk \mathcal{B}_1=1,\quad \rnk \mathcal{B}_2=2,\quad
    \rnk \mathcal{B}_3=4
    \label{eq:rankshodsm2}
\end{equation}
into Eq.~\eqref{eq:multitetrabolic} the partial multiplicity function is given by
\begin{equation}
    \beta(3)=1,\quad
\beta(1)=1, \quad
\beta(2)=
\beta(4)=0.
\end{equation}
Therefore, the degeneracy scenario in this case is an FEP with partial multiplicity $(l_{i,1},l_{i,2})=(3,1)$. 

The same scenario also occurs at $k_z=\pm\arccos(-\varepsilon)$, which follows by repeating the analysis for the corresponding Hamiltonian
\begin{equation}
    \mathcal{H}=
    -\frac{\varepsilon}{2}\left(
\begin{array}{cccc}
 0 & 0 & i & i \\
 0 & 0 & -i  & -i \\
 1 & 1  & 0 & 0 \\
 1 & 1 & 0 & 0 \\
\end{array}
\right).
\label{eq:HODSMNH2a2}
\end{equation}
We depict these FEPs in the bulk dispersion relation of Fig. \ref{hodsmbulkplot}(c), where we again set $\varepsilon=1/\sqrt{2}$.

\subsection{Non-Hermitian model 3: FEPs with\\
geometric multiplicity $\gamma_i=2$ and $(l_{i,1},l_{i,2})=(2,2)$}\label{subsec:model3}

We now turn to the second case of FEPs with geometric multiplicity $\gamma_i=2$, where the partial  multiplicities are $(l_{i,1},l_{i,2})=(2,2)$.
To obtain this scenario, we retain all expansion coefficients $q_i^{(3)}(\mathbf{k})=p_i(\mathbf{k})$ and
$r_i^{(3)}(\mathbf{k})=p_i^*(\mathbf{k})$ at their Hermitian values \eqref{eq:coefficient_a}, with the sole exception of
\begin{equation}
    q_3^{(3)}(\mathbf{k})=p_3(\mathbf{k})-\varepsilon.
    \label{eq:a32}
\end{equation}
In terms of Eq.~\eqref{eq:HODSMNHspecific}, this modification realizes NH model 3.
The dispersion relation of this case is given by
\begin{equation}
    E_{\pm,\pm}=\pm \frac{1}{\sqrt{2}} \sqrt{\cos  k_z}\sqrt{\cos  k_z\pm \sqrt{2} \varepsilon },
    \label{eq:dis3}
\end{equation}
which, upon setting to zero, gives degeneracies in the first Brillouin zone at
\begin{equation}
    k_z=\pm\pi/2 \,\,\,\, \text{and} \,\,\,\, k_z=\arccos \left(\pm\sqrt{2}\varepsilon\right).
\end{equation}
At $k_z=\pm\pi/2$, the Hamiltonian takes the form
\begin{equation}
    \mathcal{H}=\left(
\begin{array}{cccc}
 0 & 0 & -\varepsilon & 0 \\
 0 & 0 & 0  & \varepsilon \\
 0 & 0  & 0 & 0 \\
 0 & 0 & 0 & 0 \\
\end{array}
\right)\equiv h^{(3)}.
\label{eq:HODSMNH2b}
\end{equation}
We apply the recursion relation in \eqref{eq:bkHODSMNH1} to obtain the modes
\begin{equation}
    \mathcal{B}_3=\openone,
     \quad
    \mathcal{B}_2=\mathcal{H},
    \quad
    \mathcal{B}_1=\mathcal{B}_0=0,
\label{eq:bkHODSMNH2}
\end{equation}
along with their ranks,
\begin{align}
    \rnk \mathcal{B}_0=
    \rnk \mathcal{B}_1=0,\quad
    \rnk \mathcal{B}_2=2,\quad
    \rnk \mathcal{B}_3=4,
\end{align}
Using the relation \eqref{eq:multitetrabolic}, the corresponding partial multiplicity function is
\begin{equation}
    \beta(2)=2,\quad
\beta(1)=
\beta(3)=
\beta(4)=0.
\end{equation}
Therefore, as desired we encounter an FEP with partial multiplicities $(l_{i,1},l_{i,2})=(2,2)$. Since, in this case,
there are two repeated leading eigenvectors ($\beta(2)=2)$, when determining the response at this FEP one has to distinguish between the physical response strength \eqref{eq:etafromb} and the spectral response strength \eqref{eq:xifromb}.

The analogous analysis of the degeneracies at 
$k_z=\arccos\left(\pm\sqrt{2}\varepsilon\right)$,
where
\begin{equation}
    \mathcal{H}=\frac{\varepsilon}{\sqrt{2}}\left(
\begin{array}{cccc}
 0 & 0 & 1-\sqrt{2} & 1 \\
 0 & 0 & -1  & -1-\sqrt{2} \\
 1 & -1  & 0 & 0 \\
 1 & 1 & 0 & 0 \\
\end{array}
\right),
\label{eq:HODSMNH2}
\end{equation}
reveals that these are generic EPs with algebraic multiplicity 2 (this algebraic multiplicity can also be read off directly from Eq.~\eqref{eq:dis3}).
We note that these EP2s only exist in the range $-1/\sqrt{2}<\varepsilon<1/\sqrt{2}$. 

More generally,
for any fixed value of the intracell coupling parameter $t$, one can create or annihilate these EP2s by tuning the strength of the non-Hermiticity parameter $\varepsilon$ without influencing the FEPs in the bulk energy bands. 
We show a band diagrams with four generic EP2 and two FEPs in Fig. \ref{hodsmbulkplot}(d), where we set $\varepsilon=1/2$.

\subsection{Non-Hermitian model 4:\\FEPs with geometric multiplicity $\gamma_i=3$}\label{subsec:model4}
We complete our discussion of all possible degeneracy scenarios in the bulk spectrum of the non-Hermitian HODSM model by constructing a system displaying FEPs with geometric multiplicity of $\gamma_i=3$, having partial multiplicities $(l_{i,1},l_{i,2},l_{i,3})=(2,1,1)$. 

Following similar considerations as before, we set the expansion coefficients to
\begin{align}
    & q_0^{(4)}(\mathbf{k})=p_0(\mathbf{k})+\varepsilon/2, \nonumber\\
    & q_1^{(4)}(\mathbf{k})=p_1(\mathbf{k})-\varepsilon/2, \nonumber\\
    & q_2^{(4)}(\mathbf{k})= p_2(\mathbf{k})+i\varepsilon/2, \nonumber \\
    & q_3^{(4)}(\mathbf{k})= p_3(\mathbf{k}) +\varepsilon/2,
    \label{eq:a3_upper}
\end{align}
while keeping 
  $r_i^{(4)}(\mathbf{k})=p_i^*(\mathbf{k})$ fixed to the values of the Hermitian model, given in Eq.~\eqref{eq:coefficient_a}.
In terms of Eq.~\eqref{eq:HODSMNHspecific}, we obtain NH model 4.
The dispersion relation for this case is given as
\begin{equation}
    E_{\pm,\pm}=\pm\frac{1}{\sqrt{2}} \sqrt{\cos^2  k_z+(1\pm 1) \varepsilon  \cos k_z},
    \label{eq:dis4}
\end{equation}
which displays degeneracies at the locations
\begin{equation}
    k_z=\pm\pi/2,\quad \text{and} \quad k_z=\arccos(-2\varepsilon).
\end{equation}

At $k_z=\pm \pi/2$, the Hamiltonian reduces to
\begin{equation}
    \mathcal{H}=\left(
\begin{array}{cccc}
 0 & 0 & \varepsilon & 0 \\
 0 & 0 & -\varepsilon  & 0 \\
 0 & 0  & 0 & 0 \\
 0 & 0 & 0 & 0 \\
\end{array}
\right)\equiv h^{(4)},
\label{eq:HODSMNH3}
\end{equation}
leading to the modes
\begin{align}
\mathcal{B}_3=\openone,\quad
\mathcal{B}_2=\mathcal{H},\quad
\mathcal{B}_1=
\mathcal{B}_0=0,
\end{align}
with ranks
\begin{align}
    \rnk \mathcal{B}_0=
    \rnk \mathcal{B}_1=0,\quad
    \rnk \mathcal{B}_2=1,\quad
    \rnk \mathcal{B}_3=4.
\end{align}
Using again the relations \eqref{eq:multitetrabolic}, the partial multiplicity function follows as
\begin{equation}
    \beta(2)=1,\quad
\beta(1)=2,  \quad
\beta(3)=
\beta(4)=0.
\end{equation}
Therefore, at $k_z=\pm \pi/2$, the degeneracy scenario is an FEP with  
partial multiplicities  $(l_{i,1},l_{i,2},l_{i,3})=(2,1,1)$. 

In addition, following a similar calculation as discussed in the previous subsection, we determine that the degeneracy scenarios at  $k_z=\arccos(-2\varepsilon)$ are generic EP2s, which now occur for $-1/2\leq\varepsilon\leq 1/2$. We display band diagrams combining these degeneracy scenarios in Fig.~\ref{hodsmbulkplot}(e), where we set $\varepsilon=1/\sqrt{8}$.

\section{Fragmented exceptional lines formed by hinge states\label{sec:hinges}}

As a final demonstration of the richness of FEPs and their formation mechanisms, we now describe their appearance in the boundary spectrum of the HODSM models of the previous section.  In the topologically nontrivial phase of the Hermitian parent model, open boundary condition along the $x$ and $y$ directions are known to introduce hinge states localized at the four edges parallel to the $z$ axis \cite{Hodsm}.
These states appear for the parameter range $-3/2<t/s<-1/2$; otherwise, the system is gapped in the bulk, making it a trivial insulator.
The hinge states are protected by the mirror reflection symmetries of the system, and are characterized by a higher-order topological invariant, the quadrupole moment $q_{xy}$  \cite{multipole2}. 
This marks them as manifestations of the corner modes of the underlying QI model \cite{multipole1}, so that they correspond to corner states of a reduced system at a fixed $k_z$ momentum component, and in this picture form a line in momentum space which connects the two tetrabolic Dirac points.

Extending this analysis to the non-Hermitian extensions of the previous subsection, we
will see that the hinge states morph into fragmented exceptional lines, parameterized once more by the $k_z$ momentum component. As we will establish over the course of this section, the specific degeneracy structure of these states can be determined in exactly solvable configurations, while more generally it
is determined by a nonperturbative interplay of finite size effects and state deformation due to nonreciprocal non-Hermitian couplings.

\begin{figure*}
    \centering
    \includegraphics[width=0.9\linewidth]{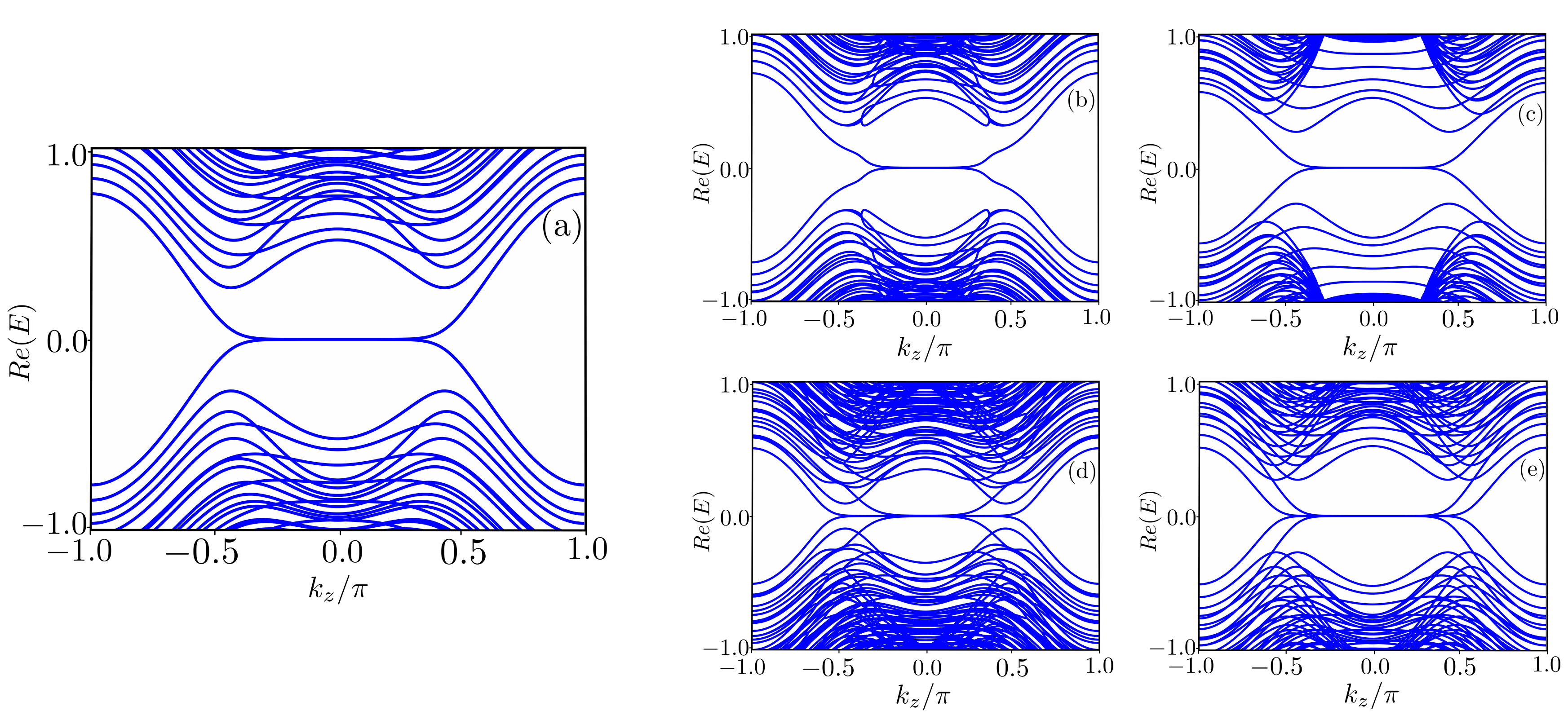}
    \caption{Energy eigenspectra of the Hermitian HODSM model and its four non-Hermitian versions for the same parameters as in Fig.~\ref{hodsmbulkplot}, but obtained for open boundary conditions in the $x$ and $y$ direction (real part only; $N_x=N_y=20$). In all cases, nearly flat four-fold degenerate hinge-state branches emerge from bulk degeneracies and extend into the momentum range around $k_z=0$. In panel (a),  the model is Hermitian, so that these states are near a tetrabolic point. In NH model 1, panel (b), the hinge states appear in the region between two generic EP4s closest to $k_z=0$ and themselves are near an EP4. In the other three models, the hinge states emerge from bulk FEPs. In NH models 2 and 3, panel (c) and (d),  they display the signatures of an FEP with partial multiplicities  $(l_{i,1},l_{i,2})=(2,2)$, while in NH model 4, panel (e), they display the signatures of an FEP with partial multiplicities  $(l_{i,1},l_{i,2},l_{i,3})=(2,1,1)$.}
    \label{hingeeigenplot}
\end{figure*}

\subsection{Model  and numerical band dispersions}

We implement open boundary conditions along the $x$ and $y$ axes, but leave the system extended into the $z$ direction (alternatively, we can consider periodic boundary conditions in this direction). In the $x-y$ plane, we enumerate the 4-site unit cells by integers $x$ and $y$ ranging from $1$ to $N_x$ and $N_y$, respectively. 
We note that both directions are equivalent, with the $Z_2$ gauge freedom allowing to place the negative-valued coupling into any of the two directions, and hence set $N_x=N_y$.
The system is then most conveniently defined and analyzed in a mixed representation given by the discrete real-space coordinates $x$ and $y$ and the continuous momentum-space variable $k_z$. 

\begin{figure*}
    \centering
    \includegraphics[width=\linewidth]{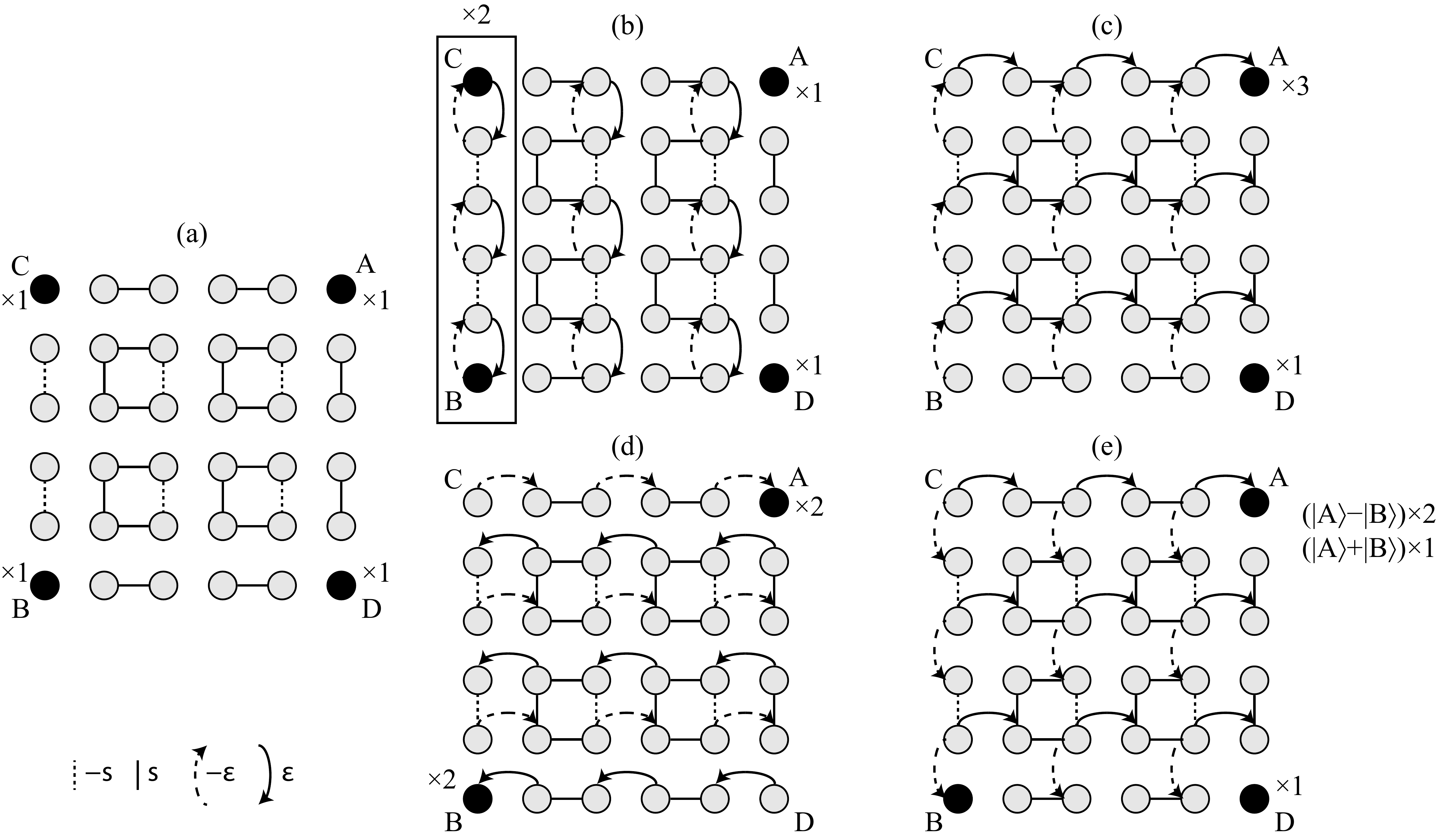}
    \caption{Hinge-state formation in the atomistic limit $s=-2t\cos k_z$ of the HODSM models, illustrated for finite systems spanning $3\times 3$ unit cells. The atomistic limit yields exactly degenerate states at zero energy, with algebraic multiplicity 4 (a,c,d,e) or algebraic multiplicity 2 (b).
    (a) In the Hermitian model, the four corner become decoupled from the remaining sites, resulting in independent zero-energy states that form a tetrabolic point [(partial multiplicities $(1,1,1,1)$]. (b) In NH model 1, the A and D corners remain decoupled and provide two independent zero-energy states that form a diabolic point with partial multiplicities $(1,1)$.
    The B and C corners join  into a non-Hermitian SSH chain (boxed in), providing two hybridized low-energy edge states for $|\varepsilon|<s$.  (c) In NH model 2, the  D corner remains decoupled, while the A corner only has incoming couplings. This results in the formation of an exact degeneracy with partial multiplicities $(3,1)$. (d) In NH model 3,  the A and B corners only have incoming couplings, and support exact zero modes with partial multiplicities $(2,2)$. (d) In NH model 4, the D corner is decoupled, and the A and B corners only have incoming couplings, leading to exact zero modes with partial multiplicities $(2,1,1)$.}
    \label{fig:atomistic}
\end{figure*}

\begin{widetext}
 This leads to the effective tight-binding Hamiltonian   
\begin{equation}
            H(k_z)=  \sum\limits_{x,y=1,1}^{N_x,N_y} \, c_{x,y}^{\dagger}\,H_0(k_z) \,c_{x,y}  
           + \left(  \sum\limits_{x,y=1,1}^{N_x-1,N_y} \,c_{x,y}^{\dagger}\,s_x\,c_{x+1,y} 
             + \sum\limits_{x,y=1,1}^{N_x,N_y-1} c_{x,y}^{\dagger}\,s_y\,c_{x,y+1}+ \text{H.c.} \right),
             \label{hingehamiltonian2}
\end{equation}
where $c_{x,y}$ is a vector of annihilation operators on the A, B, C, and D sites of the corresponding unit cell,
the intercell couplings are given by
\begin{equation}\label{seq10}
  s_x=s\left(
\begin{array}{cccc}
 0 & 0 & 1 & 0 \\
 0 & 0 & 0 & 0 \\
 0 & 0 & 0 & 0 \\
 0 & 1 & 0 & 0 \\
\end{array}
\right), \quad s_y=s\left(
\begin{array}{cccc}
 0 & 0 & 0 & 1 \\
 0 & 0 & 0 & 0 \\
 0 & -1 & 0 & 0 \\
 0 & 0 & 0 & 0 \\
\end{array}
\right),
\end{equation}
and $H_0(k_z)$ is the model-specific reduced intracell Hamiltonian.
We write this 
reduced  Hamiltonian as
\begin{equation}
   H_0(k_z)=
   \left(t +\frac{s}{2}\cos k_z\right)
    \left(
\begin{array}{cccc}
 0 & 0 & 1 & 1 \\
 0 & 0 & -1 & 1 \\
 1 & -1 & 0 & 0 \\
 1 & 1 &  0 & 0 
\end{array}
\right)
+ h
,\quad
\label{eq:hinternal}
\end{equation}
where the Hermitian case is described by 
$h=0$,
while in the four non-Hermitian extensions of the previous section
we have the additions
$h=h^{(n)}$, $n=1,2,3,4$,
precisely coinciding with the expressions given in Eq.~\eqref{eq:heps}.  
\end{widetext}

Before embarking on the detailed analysis of these models, we illustrate the emergence of low-energy hinge states in numerical energy band diagrams.
Figure \ref{hingeeigenplot} depicts numerical eigenvalue spectra of finite systems with $N_x=N_y=20$ as a function of $k_z$, where the coupling parameters  $s=-t=1$ and strength $\varepsilon$ of the non-Hermitian modifications are the same as in the bulk analysis (cf.~Fig.~\ref{hodsmbulkplot}). 
In the Hermitian system the hinge states appear as almost-degenerate states near zero energy in the $k_z$ range between the two Dirac points. In the non-Hermitian 
models, analogous states emanate from the corresponding bulk degeneracy points, so that at $k_z=0$ there are always four almost-degenerate states that approach exact degeneracy in the limit $N_x, \, N_y\rightarrow\infty$. 
The residual  splitting of these states is due to finite size effects, whose crucial role for the achieved degeneracy structures we discuss further below.
We now set out to determine the degeneracy patterns achieved in this way, for which we address different scenarios.

\subsection{Atomistic limit}
Inspecting the internal Hamiltonian \eqref{eq:hinternal} we see that it reduces to $H_0(0)=h$ if $s\cos k_z=-2t$, which is realized, for instance, for $k_z=0$ and $s=-2t$. In the Hermitian case ($h=0$), this defines the atomistic limit of the associated QI.
As shown in Fig.~\ref{fig:atomistic}(a),
the four corners at $(x,y)=(1,1)$, $(N_x,1)$, $(1,N_y)$, and  $(N_x,N_y)$ decouple from the rest of the system and support four fully localized corner states with exact energy $E=0$. Therefore, these states form a tetrabolic point with partial multiplicities $(1,1,1,1)$. 
For the discussion of the modification of this degeneracy in the non-Hermitian extensions we refer to the corners according to the label A, B, C, or D of the terminating site. We focus on the description of the right eigenstates, noting that analogous arguments can be put forward for the left eigenstates. The corresponding formation mechanisms are illustrated in Fig.~\ref{fig:atomistic}(b-e). 

In NH model 1 the additional couplings of strength $\varepsilon$ do not connect to the A and D corners, resulting in two exact zero-energy states that then form a diabolic point with partial multiplicities $(1,1)$.  The B and C corners become part of a subsystem that remains uncoupled from the remaining sites, and takes the form of a non-Hermitian Su-Schrieffer-Heeger (NHSSH) chain, originally introduced in the study of the non-Hermitian skin effect (NHSE) \cite{Yao2018}. We focus on the range $|\varepsilon|<|s|$, where this chain is in the topologically nontrivial configuration. 
The NHSSH
chain then supports two hybridized edge states with a small but finite energy that approaches zero as the system size is increased. 
In the limit of a large system, the four low-energy states in the system then approach a tetrabolic point with partial multiplicities $(1,1,1,1)$. We note that at this specific parameter point, the NHSE does not apply, as the nonreciprocal couplings are still of exactly the same strength. In the presence of the NHSE, the two NHSSH states would be distorted to be localized at the same end, leading to a $(2,1,1)$ degeneracy structure. We will encounter this mechanisms in the discussion of generic parameter choices below, where it is then further modified by the symmetry-constrained hybridization of all states. 

In the remaining three non-Hermitian models,
the non-Hermitian modifications preserve the exact fourfold algebraic degeneracy of the corner states.
These states then realize the same partial multiplicity patterns as we observed for the bulk Dirac points in these three models. In model 2, the $(3,1)$ partial multiplicity structure is realized by corner states localized in the A and D corners. Here, the A corner is distinguished by the fact that it only has an incoming coupling, but no outgoing couplings. A state localized on this site then serves as an exact right eigenstate with zero energy.
We can verify that this state has partial multiplicity three by direct modal analysis, yielding the same ranks as stipulated for the bulk Dirac point of this model in Eq.~\eqref{eq:rankshodsm2}, and this is also supported by numerical computations.
In model 3, corners A and B have only incoming couplings and thereby support exact zero-energy states. These two corners reside in two mutually decoupled subsystems, each of which is associated with a partial multiplicity of 2. In model 4, the D corner is decoupled and provides one independent zero-energy state. The A and B corners again have only incoming couplings, but these now originate from a common bulk. From the modal analysis, we obtain the overall partial multiplicity structure $(2,1,1)$, where numerical construction of the eigenstates reveals that the two-fold multiplicity arises from the antisymmetric combination $|A\rangle-|B\rangle$ of the corner states.

\subsection{Double-semi-infinite geometry}
To facilitate the discussion of hinge states for generic parameters away from the atomistic limit, 
we next consider their formation in double-semi-infinite systems.
This refers to the geometry where the system has a single corner, such as at $(x,y)=(1,1)$, from which it extends infinitely into the positive $x$ and $y$ directions (as well as infinitely into the $z$ direction). In these systems, there is only a single hinge state, and hence no degeneracy. However, this state is an exact zero-energy mode and can be constructed analytically. This reveals the importance of nonreciprocal couplings and the connection to the NHSE, which we will exploit later when we discuss the hybridization of several such states in finite systems.

In the bulk description, we analyzed the HODSM systems for real momentum values
$(k_x,k_y,k_z)$. Hinge and corner states have the physical character of confined states that decay exponentially into the bulk of system.  
The states then correspond to complex values of $k_x$ and $k_y$, where the imaginary parts determine the decay rates into the system. 
For the Hermitian parent model, the corresponding analytical continuation of the Bloch Hamiltonian \eqref{Hodsm2} takes the form
\begin{equation}
    \mathcal{H}(\mathbf{k})=\left(
\begin{array}{ccc}
 0 & P(\mathbf{k})  \\
{} [P(\mathbf{k}^*)]^\dagger & 0 
\end{array}
\right),
\label{Hodsm2complex}
\end{equation}
which for complex momenta is generally non-Hermitian. Zero-energy modes can then be found for
\begin{equation}
\det P(\mathbf{k}) = 0 \quad \mbox{or} \quad \det  P(\mathbf{k}^*) = 0.
\end{equation}

For the non-Hermitian modifications based on Eq.~\eqref{eq:HODSMNH}, the corresponding zero-mode quantization condition takes the form
\begin{equation}
\det Q(\mathbf{k}) = 0 \quad\mbox{or}\quad\det  R(\mathbf{k}) = 0
\label{eq:nonhermitianquantapp},
\end{equation}
hence, is formally the same as in Eq. \eqref{eq:nonhermitianquant}. However, in the specification of the expansion coefficients $r_i^{(n)}$ of the four different models we have to replace $p_i^*(\mathbf{k})$ by $[p_i(\mathbf{k}^*)]^*$.

The stated quantization conditions can be made manifest by considering the states confined to the $(x,y)=(1,1)$ corner of the system. 
Inspecting the boundaries of the system with the help of Fig.~\ref{fig:QI}, the boundary conditions imply vanishing of the amplitudes on the A and D sites in the beyond-the-boundary unit cells with $x=0$, 
and vanishing of the amplitudes on the A and C sites in the beyond-the-boundary unit cells with $y=0$. These patterns are then maintained throughout the lattice. Therefore we seek zero-mode solutions confined to the B sublattice. Applying a Bloch vector to the Hamiltonian  \eqref{Hodsm2complex} of the Hermitian system, we read off that this requires 
\begin{equation}
P_{21}(\mathbf{k}^*)=P_{22}(\mathbf{k}^*)=0.
\end{equation}
At a given real $k_z$ this condition is fulfilled for 
\begin{equation}
k_x=k_y=-i \ln \left(-\frac{t}{s}-\frac{1}{2}\cos k_z\right)\equiv -i q_0.
\end{equation}
This describes states that decay into the bulk if $|t/s+(1/2)\cos k_z|<1$, which covers the region between the two Dirac points specified in Eq.~\eqref{eq:hermitiandiracpoints1}.
Analogously, we obtain states confined to the three other sublattices when we consider double-semi-infinite systems anchored at the other corners.

In the non-Hermitian models, the rotational symmetry is generally broken, so that each corner has to be studied separately. 
Applying the same principles as in the Hermitian case to the $(x,y)=(1,1)$ corner in NH model 1, we obtain a state localized on the B sublattice for which the decay lengths are determined by 
\begin{align}
k_x&=-i q_0,
\\
k_y&= -iq_\varepsilon.
\end{align}
where we define
\begin{equation}
q_\varepsilon =\ln \left(-\frac{t}{s}-\frac{1}{2}\cos k_z-\frac{\varepsilon}{s}\right).
\label{eq:qeps}
\end{equation}
We note that the decay length in the $x$ direction remains unchanged, but the decay length in the $y$ direction is modified. For our standard choices of parameters $s=-t=1$ and positive $\varepsilon$, the hinge state on the B sublattice becomes more confined to its supporting corner.
Conversely, we find that the confinement of the hinge state on the $C$ sublattice reduces in $y$ direction according to a complex wavenumber $k_y=iq_{-\varepsilon}$, while the confinement of the hinge states associated with the $A$ and $D$ sublattices remains unchanged. 

\begin{table}[t]
    \centering
    \begin{tabular}{ccccccccc}
    \hline\hline
        NH model  &  \multicolumn{2}{c}{A corner}
         &  \multicolumn{2}{c}{B corner}
         &  \multicolumn{2}{c}{C corner}
         &  \multicolumn{2}{c}{D corner}
         \\
         & $k_x$ & $k_y$ 
         & $k_x$ & $k_y$ 
         & $k_x$ & $k_y$ 
         & $k_x$ & $k_y$ 
         \\
         \hline
        1 & $iq_0$ & $iq_0$
        & $-iq_0$ & $-iq_\varepsilon$
        & $-iq_0$ & $iq_{-\varepsilon}$
        & $iq_0$ & $-iq_0$
         \\
        2 & $iq_0$ & $iq_0$
        & $-iq_0$ & $-iq_\varepsilon$
        & $-iq_\varepsilon$ & $iq_0$
        & $iq_0$ & $-iq_0$
        \\
        3 & $iq_0$ & $iq_0$
        & $-iq_0$ & $-iq_0$
        & $-iq_{-\varepsilon}$ & $iq_0$
        & $iq_{\varepsilon}$ & $-iq_0$
        \\
        4 & $iq_0$ & $iq_0$
        & $-iq_0$ & $-iq_0$
        & $-iq_{\varepsilon}$ & $iq_{\varepsilon}$
        & $iq_0$ & $-iq_0$
        \\
        \hline\hline
    \end{tabular}
    \caption{Complex wavenumbers describing the decay of hinge states in double-semi-infinite geometries anchored at the A, B, C, or D corner of the for non-Hermitian HODSM models. The quantity $q_\varepsilon$ is defined in Eq.~\eqref{eq:qeps}. Whenever it appears, the localization of the state towards the corner into the specified spatial direction is enhanced, while the appearance
    of $q_{-\varepsilon}$ signifies a reduced localization. These results apply to the right eigenstates.   
    }
    \label{tab1}
\end{table}

These findings, along with their corresponding forms for the other three non-Hermitian models, are summarized in Table \ref{tab1}. According to this, in NH model 2, the localization of the B hinge state is enhanced in the $y$ direction, while  the localization of the C hinge state is enhanced in the $x$ direction.
In NH model 3, the localization of the C hinge state is reduced in the $x$ direction, while  the localization of the D hinge state is enhanced in the $x$ direction.
In NH model 4, the localization of the C hinge state is enhanced in the $x$ and $y$ directions.

In a finite system the states hybridize nontrivially. This generically results in finite energy splittings, so that the states no longer constitute exact zero modes; however, these splittings decrease rapidly as the system size increases. In the non-Hermitian models, a much more drastic effect occurs due to the NHSE, which arises from interplay of hybridization and nonreciprocal couplings. According to this effect, the enhanced localization of a state in a given direction results in an enhanced weight that grows exponentially with system size, while the weight of the state localized into the `opposite direction' becomes  exponentially suppressed. Here, the C corner is opposite the B corner in $y$ direction, and opposite the A corner in $x$ direction, while the
D corner is opposite the A corner in $y$ direction, and opposite the B corner in $x$ direction.
Before we can establish this in detail, we have to account for the symmetries of the different models.

\subsection{Interplay with symmetries}

The detailed hybridization rules for the energy splittings and spatial localization patterns are constrained by symmetries of the various models, which we describe next. These are the Kramers degeneracy of the Hermitian model, a generalized reflection symmetry that also applies to NH model 4,
as well as a generalized transposition symmetry and sublattice sum rule and that are particularly relevant to NH model 2. 
All these symmetries and constraints hold at finite system sizes with open boundary conditions, where the hybridization generically results in finite energy splittings.

\subsubsection{Kramers degeneracy}

Even after hybridization, all states in the Hermitian system, including the hinge states, display an exact even-fold degeneracy.  This degeneracy is enforced by the interplay of time-reversal symmetry and the $C_4$ rotational symmetry specified in Eq.~\eqref{eq:c4}. Given an eigenstate $\mathbf{u}$, we obtain a second eigenstate $\mathbf{u}'=C_4^2\mathbf{u}^*$
which has the same energy but is orthogonal $\mathbf{u}$. To see this, we first note that as $\mathcal{K}C_4^2$ commutes with the Hamiltonian, where $\mathcal{K}$ again denotes complex conjugation, $\mathbf{u}'$ is also an eigenstate.
To show that the two states are orthogonal to each other, we 
 utilize  $C_4^4=-\openone$ to 
evaluate their scalar product as
\begin{equation}
\langle \mathbf{u}|\mathbf{u}'\rangle=
\langle \mathbf{u}|C_4^2\mathbf{u}^*\rangle
=
\langle C_4^2\mathbf{u}|C_4^4\mathbf{u}^*\rangle
=
-\langle \mathbf{u}^{\prime*}|\mathbf{u}^*\rangle
=
-\langle \mathbf{u}|\mathbf{u}'\rangle
\label{eq:kramersproof}
\end{equation}
Therefore, this scalar product  vanishes. 
We recognize this feature as a specific variant of Kramer's degeneracy, induced here by a generalized antiunitary time-reversal symmetry $\tilde{\mathcal{T}}=\mathcal{K} C_4^2$ that fulfills $(\tilde{\mathcal{T}})^2=-\openone$. 

Therefore, generically, the corner states in the Hermitian system hybridize to form two exactly degenerate diabolic points with opposite energies $\pm E_c$. These energies approach $E_c\to 0$ as the system size is increased. In this limit, the states can then be viewed as a symmetry-constrained unfolding of a tetrabolic point.

\subsubsection{Generalized reflection symmetry}
We observe that the presence of the $\pi$ fluxes in the models gives rise to a generalized reflection symmetry $\tilde R_a=Z_CR_a$ about the antidiagonal. Here, $R_a$ denotes conventional reflection of the lattice about the antidiagonal, which maps, for instance, the A corner onto the B corner. In the Hermitian model, the reflection also maps some positive couplings onto negative couplings. This is exactly compensated by the gauge transformation $Z_C$, defined to multiply all amplitudes on the C sublattice by $-1$. Inspecting all non-Hermitian models, we see that this symmetry is also satisfied for NH model 4.

The significance of this symmetry is as follows. In the presence of an ordinary reflection symmetry, we would expect to obtain three symmetric hinge states and one antisymmetric hinge state (in the atomistic limit, the symmetric states would be $|A\rangle+|B\rangle$, $|C\rangle$, and $|D\rangle$, while the antisymmetric states would be $|A\rangle-|B\rangle$). However, under the generalized reflection symmetric the $|C\rangle$ state becomes antisymmetric. The hybridization of the states in NH model 4 has to respect this modification, so that under this symmetry we expect to obtain two symmetric and two antisymmetric hinge states.

\subsubsection{Generalized transposition symmetry in NH model 2}

In NH model 2, a variant of Kramers degeneracy can be constructed when we replace the antilinear time-reversal symmetry $\tilde{\mathcal{T}}$ by the matrix transposition, which we will denote as $\mathcal{T}'$ \cite{schomerus2013from}. The distinction of these two operations, which coincide for Hermitian systems, is at the heart of the general classification of non-Hermitian universality classes \cite{kawabata2019symmetry}.
We then find that the operator $\tilde R_a\mathcal{T}'$ commutes with the Hamiltonian,  where $\tilde R_a$ is the generalized reflection operation of the system about the antidiagonal described above.
We can again use this symmetry to construct for each eigenstate $\mathbf{u}$ a partner state $\mathbf{v}=(R_a\mathbf{u})^T$, where we must treat $\mathbf{u}$ as a right eigenstate and
$\mathbf{v}$ as a left eigenstate. As we will see next, in NH model 2 these two states are furthermore orthogonal to each other. In contrast to the Kramers degeneracy in Hermitian systems, this then signifies exceptional points throughout the entire energy spectrum.

\subsubsection{Sublattice sum rules}
In the discussion of the bulk dispersion relation
\eqref{eq:trace_cond},
we mentioned that the HODSM can be useful interpreted as a 
square-root topological insulator \cite{Arkinstall2017}. 
We now apply this approach to the finite systems to identify a nontrivial sum rule.
To formulate this rule, we interpret the Hamiltonian in terms of blocks $H_{XY}$, where $X,Y\in(A,B,C,D)$ labels the sublattice. Due to the chiral symmetry, the blocks $H_{XY}$ with $X,Y\in(A,B)$ 
or  $X,Y\in(C,D)$ vanish. However, an additional feature emerges in the Hermitian model, where the squared Hamiltonian $H^2$ takes a further simplified structure, in which only the four diagonal blocks $(H^2)_{AA}$, $(H^2)_{BB}$, $(H^2)_{CC}$,  and $(H^2)_{DD}$ are finite. This results because of additional sublattice sum rules such as 
\begin{equation}
(H^2)_{BA}=H_{BC}H_{CA}+H_{BD}H_{DA}=0.
\label{eq:sumrule1}
\end{equation}

The  significance of these sum rules is that they allow us to construct exact eigenstates that vanish on a given sublattice. In particular, the specific sum rule \eqref{eq:sumrule1} provides right eigenstates that vanishes on the B sublattice, as well as left eigenstates that vanishes on the A sublattice; and these states are always orthogonal to each other, even if they belong to the same energy. The right eigenstates $\mathbf{u}_l=(\mathbf{a}_l^T,\mathbf{b}_l^T,\mathbf{c}_l^T,\mathbf{d}_l^T)^T$ are found from the reduced eigenvalue problem
\begin{equation}
(H^2)_{AA}\mathbf{a}_l=E_l^2\mathbf{a}_l
\label{eq:redH1}
\end{equation}
for the  components $\mathbf{a}_l$ on the A sublattice, which then determine the components on the other sublattices as $\mathbf{b}_l=0$,
\begin{align}
\mathbf{c}_l=\frac{1}{E_l}H_{CA}\mathbf{a}_l,
\\
\mathbf{d}=\frac{1}{E_l}H_{DA}\mathbf{a}_l.
\end{align}
Analogously, the left eigenstates 
$\mathbf{v}_m=(\mathbf{a}'_m,\mathbf{b}'_m,\mathbf{c}'_m,\mathbf{d}'_m)$
are  obtained from the reduced eigenvalue problem
\begin{equation}
\mathbf{b}_m'(H^2)_{BB}=E_m^2\mathbf{b}_m'
\label{eq:redH2}
\end{equation}
for the components $\mathbf{b}'_m$ on the B sublattice, while the components on the other sublattices are then given by $\mathbf{a}'_m=0$,
\begin{align}
\mathbf{c}'_m=\frac{1}{E_m}\mathbf{b}'_mH_{BC},
\\
\mathbf{d}_m'=\frac{1}{E_m}\mathbf{b_m}'H_{BD}.
\end{align}
The consistency and orthogonality of these two states can be directly confirmed from the sum rule \eqref{eq:sumrule1}.

An intriguing situation arises when the two reduced eigenvalue problems \eqref{eq:redH1} and  \eqref{eq:redH2} produce the same eigenvalues. 
This is the case in the Hermitian model. As in this case, the right and left eigenstates are in strict correspondence with each other, this reasoning amounts to the explicit construction of two independent states for any energy eigenvalue, without resorting to the Kramer's degeneracy derived in Eq.~\eqref{eq:kramersproof}. Furthermore, in this model, analogous solutions can be constructed that vanish on any of the other three sublattices.

In the non-Hermitian models, the underlying sum rules are violated except for
NH models 2 and 4. In NH model 2, the specific sum rule \eqref{eq:sumrule1}  applies, while in NH model 4 we observe the sum rule $(H^2)_{CD}=0$ (the analogous sum rules on the other sublattices do not apply in these models).
Furthermore, in NH model 2, the generalized transposition symmetry enforces the same spectrum for the reduced 
eigenvalue problems \eqref{eq:redH1} and  \eqref{eq:redH2}.
However, now the orthogonality of the constructed right and left eigenstates no longer signifies a diabolic point. Instead, generically, this feature has to be interpreted as the self-orthogonality of states at an exceptional point. Consequentially, in NH model 2, all states, including the hinge states, have evenfold algebraic degeneracy, and generically are exceptional points.

\subsection{Degeneracy patterns for generic parameters\label{sec:hingeformation}}

With these preparations in place, we now proceed to describe the FEP degeneracy patterns of the hinge states across the different models for generic parameters away from the atomistic limit, highlighting not only the patterns themselves but also the underlying mechanisms that give rise to their intricate structure.

\begin{figure}[t]
\centering
\includegraphics[width=\linewidth]{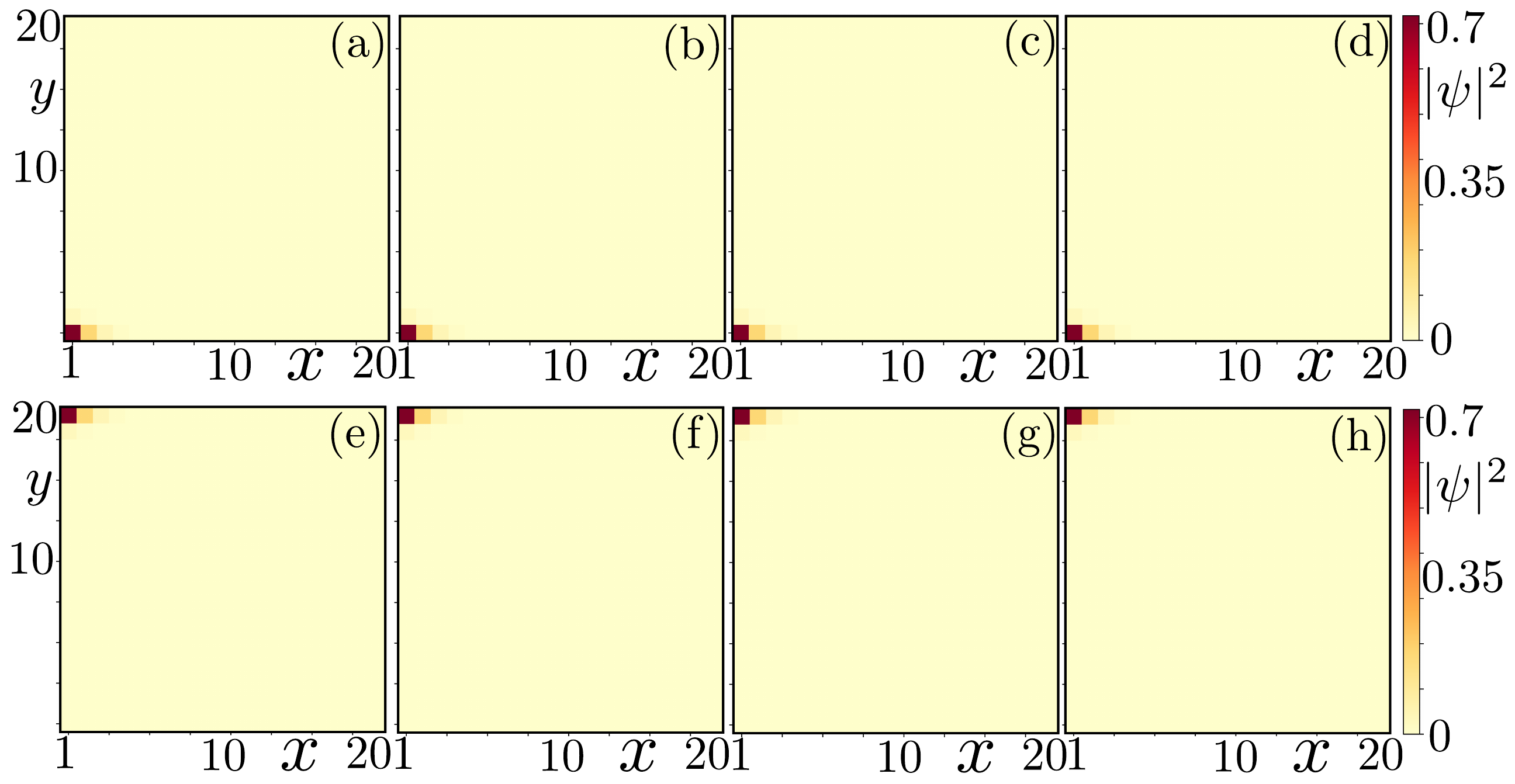}
\caption{Density plots of the intensity in each unit cells of the (a)-(d) right hinge eigenstates and (e-f) left hinge eigenstates in NH model 1, evaluated at $k_z=0$ in the band structure of Fig.~\ref{hingeeigenplot}(b), where $s=-t=1$ and $\varepsilon=1/\sqrt{2}$. The system spans 20 unit cells in the $x$ and $y$ directions, and the intensities are summed over the four sites in each unit cell. All four right eigenstates are localized in the lower left corner,
while four left eigenstates are localized in the upper left corner, placing the system into the vicinity of an EP4. 
As explained in the text,  we can attribute this pattern to the nonperturbative influence of the non-Hermitian skin effect on these states.
}
\label{hingesgm1}
\end{figure}

\begin{figure}[t]
\centering
\includegraphics[width=\linewidth]{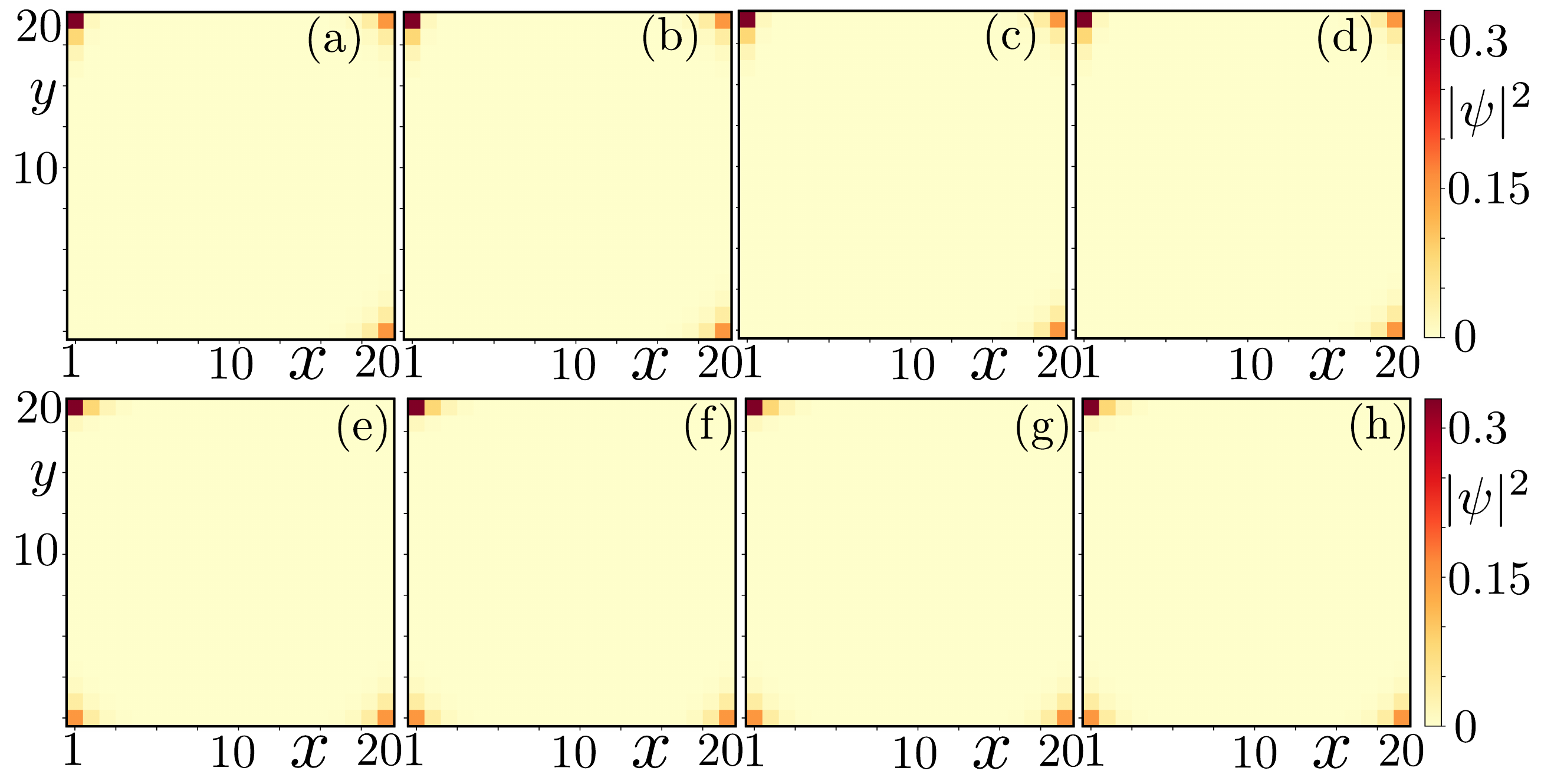}
\caption{Intensity plots as in 
Fig.~\ref{hingesgm1} but for NH model 2, corresponding to $k_z=0$ in Fig.~\ref{hingeeigenplot}(c). In this case, the right eigenstates tend to localize on the C sublattice, while the intensity in the B sublattice vanishes due to the sublattice sum rule explained in Sec.~\ref{sec:hingeformation}, which furthermore enforces the hinge states to pair up into two EP2s. This brings the states into the vicinity of an FEP with partial multiplicities $(l_{i,1},l_{i,2})=(2,2)$.
}
\label{hingesgm2}
\end{figure}

\begin{figure}[t]
\centering
\includegraphics[width=\linewidth]{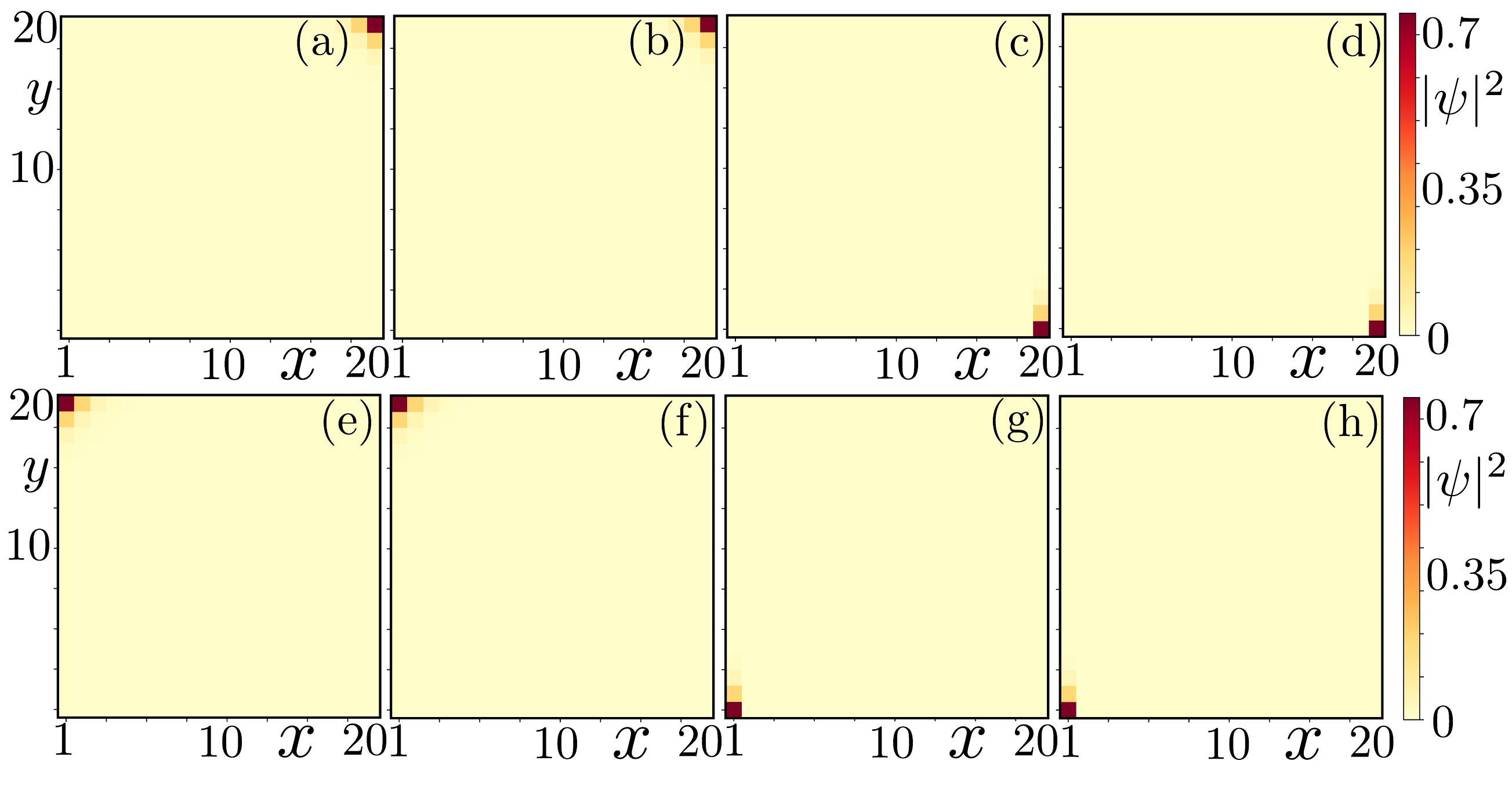}
\caption{Intensity plots as in 
Fig.~\ref{hingesgm1} but for NH model 3, corresponding to $k_z=0$ in Fig.~\ref{hingeeigenplot}(d), where $\varepsilon=1/2$.
Due to the NHSE, two of the right eigenstates are localized in the upper right corner, while the other two are localized in the lower right corner,  placing the system into the vicinity of an FEP with $(l_{i,1},l_{i,2})=(2,2)$. 
}
\label{hingesgm3}
\end{figure}

\begin{figure}[t]
\centering
\includegraphics[width=\linewidth]{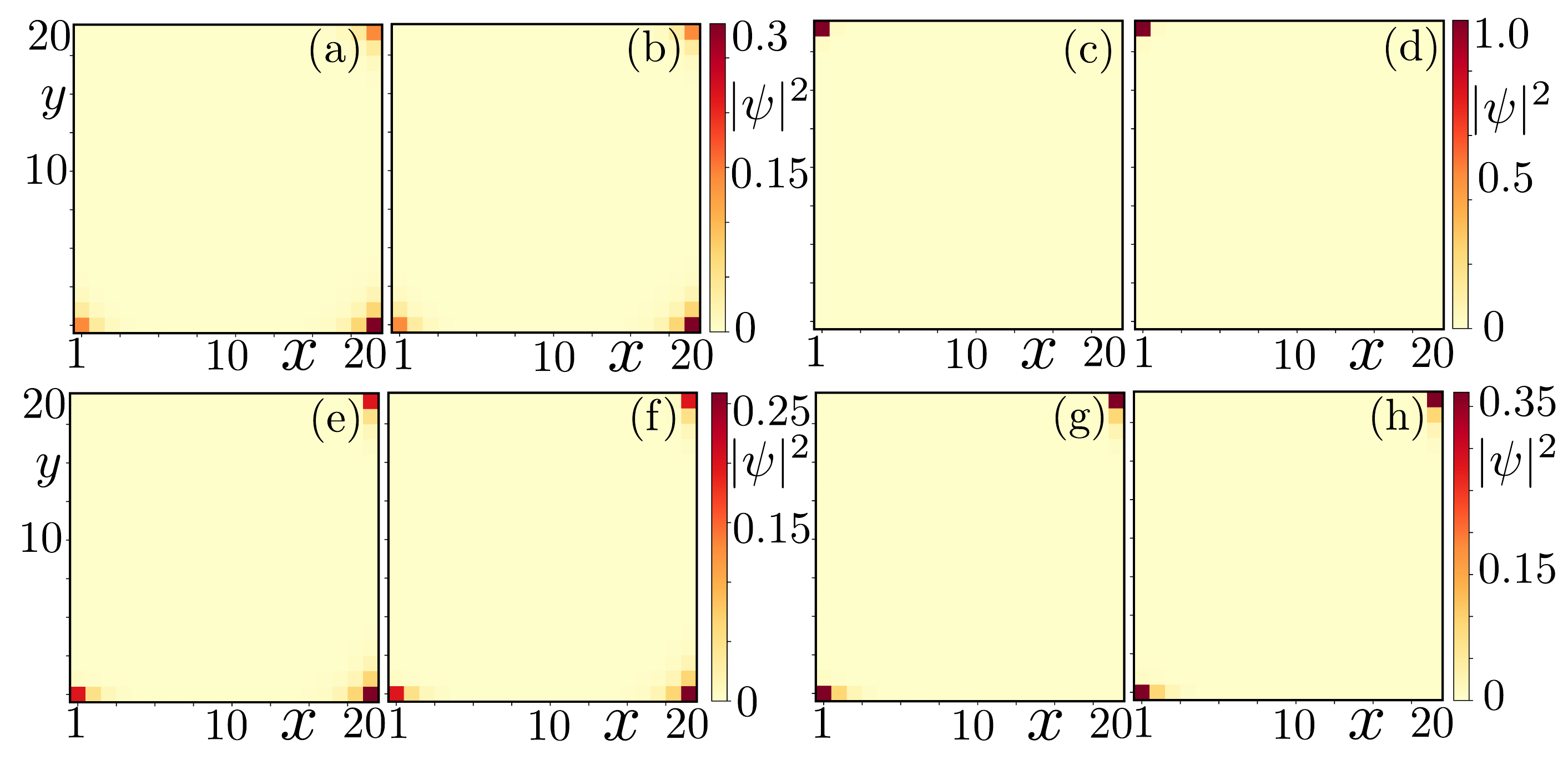}
\caption{Intensity plots as in 
Fig.~\ref{hingesgm1} but for NH model 4, corresponding to $k_z=0$ in Fig.~\ref{hingeeigenplot}(e), where $\varepsilon=1/\sqrt{8}$. For two of the right eigenstates, the NHSE enhances the intensity component in the upper left corner, resulting in two almost identical states shown in panels (c) and (d).
These states transform antisymmetrically under the generalized reflection symmetry of this model, leaving two independent symmetric right states shown in panels (a,b). This places the system near an FEP with $(l_{i,1},l_{i,2},l_{i,3})=(2,1,1)$. 
}
\label{hingesgm4}
\end{figure}

To establish these general patterns, we first revisit Fig.~\ref{hingeeigenplot} depicting the numerical eigenvalue spectra of finite systems with  $s=-t=1$ and  $N_x=N_y=20$ as a function of $k_z$.
Numerically, we can utilize the overlaps of the states on the hinge branches to infer which degeneracy scenario is then approached. This places the hinge states in NH model 1 near an EP4, in non-Hermitian models 2 and 3 near an FEP with partial multiplicities $(l_{i,1},l_{i,2})=(2,2)$, and in model 4 near a  FEP with partial multiplicities $(l_{i,1},l_{i,2},l_{i,3})=(2,1,1)$. 
In models 1 and 2, these multiplicities differ from those in the bulk and atomistic scenarios.  
Furthermore, as depicted in 
Figs.~\ref{hingesgm1}, \ref{hingesgm2}, \ref{hingesgm3}, and \ref{hingesgm4}, in all non-Hermitian models, the spatial intensity distributions of the states differ noticeably from those in the atomistic limit (cf. Fig. \ref{fig:atomistic}).


The modified degeneracy patterns and spatial distributions arise from a highly nontrivial interplay of the non-Hermiticity skin effect and non-Hermitian symmetries in these finite systems described above. The hinge states then hybridize according to the following mechanisms (we again focus on the right eigenstates).

In model 1 (Fig.~\ref{hingesgm1}), the NHSE enhances the corner-state component on the B sublattice. Since all symmetries are broken in this model, all states inherit this enhancement. This enhancement increases exponentially with system size, so that all states obtain a very large overlap, as characteristic for the vicinity of an EP4. 

In model 2 (Fig.~\ref{hingesgm2}), the NHSE enhances corner state components both on the C and B sublattices. However, due to the 
described B lattice sum rule,
the hybridized solutions do not involve the B sublattice. Therefore, the hybridized states only show an enhancement on the C sublattice. Furthermore, as described above, in this model, Kramers degeneracy becomes replaced by a sum-rule symmetry that protects exceptional points. Therefore, the low-energy hinge states appear as two exact EP2s that approach each other as the system size increases.
This is then consistent with the vicinity of an FEP with partial multiplicities $(2,2)$. Indeed, the $(3,1)$ multiplicity pattern observed in the atomistic case is only compatible with the chiral symmetry observed by all models when energies exactly vanish.

In model 3 (Fig.~\ref{hingesgm3}), the NHSE reduces the spatial localization of the C component and enhances the spatial localization of the D component. These enhancements are both in the $x$ direction and systematically shift the weights onto the A and D sublattices. This leads to the formation of two states in each corner whose overlap increases with system size, signifying again the vicinity of an FEP with partial multiplicities $(2,2)$.  

In model 4  (Fig.~\ref{hingesgm4}), the NHSE enhances the C component both in $x$ and $y$ direction. Furthermore, the system obeys the stated generalized reflection symmetry $\tilde R_a$ about the antidiagonal. In the Hermitian case, the corner states can be combined into two symmetric and two antisymmetric states with respect to this generalized symmetry. Due to the NHSE on the C component, which transforms antisymmetrically, the two antisymmetric states acquire a large overlap when the system size is large. This then results in the observed vicinity of an FEP with partial multiplicities $(2,1,1)$.

These intricate mechanisms and features further
underline the structural and phenomenological richness of the fragmented exceptional points described in this work.

\section{Summary and conclusions}
\label{sec:conclusion}
In summary, we have introduced a systematic and efficient framework to identify, classify, and construct non-Hermitian systems exhibiting fragmented exceptional points,   complex degeneracy scenarios extending the notion of ordinary exceptional points to situations in which the eigenvectors are only partially degenerate.
Our framework directly starts from insights into the physical signatures of these degeneracies, and delivers concrete conditions for their formation that can be efficiently evaluated in specific models. We demonstrated this by constructing a wide range of FEPs in two lattice models, based on the Lieb lattice and a higher-order Dirac semimetal.

Fragmented exceptional points occur naturally in the parameter space of all non-Hermitian models, and complete the conceptual description of this broad class of systems.
They venture into uncharted territory in the study of non-Hermitian systems, where one benefits from their large degree of variety and complex spectral interplay, enhanced by the rapid proliferation of partial multiplicity configurations with increasing algebraic multiplicity. 
Nevertheless, we could demonstrate that this complexity is naturally captured by a single unifying object, the modes of the adjugate matrix, which can be efficiently evaluated in specific models, and moreover directly determine the signatures of these degeneracies in the physical and spectral response to external driving, quantum noise, and parameter variations. These firm links are embedded in
Eqs.~\eqref{eq:superlorentzian} and \eqref{eq:powerlawresponse}, which specify the qualitative nature of the spectral and physical response in the form of powerlaw response functions and super-Lorentzian resonance lineshapes, 
Eqs.~\eqref{eq:etafromb} and ~\eqref{eq:xifromb},  which extract the spectral and physical response strengths that quantify these physical signatures of the system, and the central Eq.~\eqref{eq:betalresult}, which resolves the partial multiplicities of the degeneracy in terms of the same algebraic quantities.

In the application to the specific models, we encountered an intricate interplay of non-Hermitian effects and symmetry constraints, which was further accentuated for FEPs arising from topologically protected hinge states. 
The models that we studied are of direct physical significance. 
Lieb lattice models have been realized on a wide range of platforms, such as two-dimensional optical waveguide arrays fabricated using the femtosecond laser writing techniques  \cite{photoniclieb1,photoniclieb2}, cold atomic gases \cite{lieb3}, microwave resonator arrays \cite{Poli_2017}, and
polariton exciton systems \cite{Whittaker2018}. With our complete classification of the bulk degeneracies in these systems, they can now be designed to demonstrate the minimal version of an FEP, where three algebraically degenerate modes support two linearly independent eigenstates. The implementation of this scenario is further facilitated by our demonstration of how this can be achieved in reciprocal settings, leading to a model in which the FEPs can be studied in detail on an exceptional surface.  We anticipate that more complex FEPs can be designed in Lieb lattices with larger unit cells, such as described in Ref.~\cite{bpal}. 
Within the present study, we realized such more complex scenarios based on a three-dimensional model of a higher-order Dirac semimetal, which supports degeneracies up to fourth order. Experimental platforms realizing the non-Hermitian variants that we proposed are again readily available, for instance, in the form of acoustic topological insulators \cite{Zhang2021} and topolectric circuits \cite{Zou2021}.
In particular, these platforms also allow to realize nonreciprocal non-Hermitian systems with open boundary conditions supporting topological hinge and corner states, allowing to study the intricate formation mechanisms of hinge FEPs in the proposed models in a concrete physical setting.

Prospective applications of FEPs include sensors operating on exceptional surfaces, modifying the detection principles of exceptional-point sensors so that the vicinity of an FEP results in a drastic change of the spectral and physical response. 
The occurrence of FEPs at certain points on an exceptional surface also opens up a wide range of new possibilities for fundamental studies, including those concerning the structure of the Riemann surfaces around these special points. Such efforts would aim to extend the well-established classifications for generic EPs \cite{ryu1,ryu2,Ryu2024} to the unexplored domain of FEPs. Our formalism opens a realistic avenue for such studies as it efficiently circumvents the technical problems of the Arnold-Jordan normal form, which breaks down precisely for these scenarios.
More generally, the key results presented in this work are platform-independent, and the framework freely applies to any model based on a finite-dimensional effective Hamiltonian matrix, covering an overwhelming range of non-Hermitian models and systems. Given these features, the presented formalism and design principles complete the description of non-Hermitian degeneracies conceptually and physically and open a broad new frontier of non-Hermitian physics.

The data that support the findings of this work are openly
available \footnote{Research datasets at \url{https://doi.org/10.5281/zenodo.16575524}}.

\begin{acknowledgments}
This research was funded by EPSRC via Grant No. EP/W524438/1.
\end{acknowledgments}


\appendix

\section{Detailed derivation of the algebraic characterization of FEPs\label{app:rankcondition}}
In this Appendix, we present the detailed 
derivation of the algebraic characterization of FEPs of Sec.~\ref{sec:pepconditions}.
We recall that this is based on introducing the shifted Hamiltonian $A=H-E_i\openone$, from which the modes $\mathcal{B}_k$ are obtained via the Faddeev-LeVierre recursion relation \eqref{eq:flv}. The following combines insights from Ref.~\cite{bid2024uniform} with a final step in which we resolve the complete geometric partial multiplicities $\beta_i,l$ from the rank of the modes.

As in Ref.~\cite{bid2024uniform}, 
we start with the simple observation that
the leading coefficients 
\begin{equation}
    c_k=0, \quad k=0,1,2,\ldots,\alpha_{i}-1 
    \label{eq:ckconditionAPP}
\end{equation}
of the shifted characteristic polynomial \eqref{eq:shiftedq} vanish according to the algebraic multiplicity $\alpha_i$.
With the relation \eqref{eq:ckfromb}, this is cast into the form of Eq.~\eqref{eq:ckcondition} in the main text.

In terms of this shifted eigenvalue problem, the geometric multiplicity $\gamma_i$ 
is given by the number of independent solutions of the eigenvalue equation
\begin{equation}
\label{eq:evecshifted}
A\mathbf{u}_i=0.
\end{equation}
We can again determine this number purely algebraically, by utilizing the determinantal minors 
$M_{[I],[J]}^{(k)}(A)=\det(A_{[I],[J]})$
of $A$.
These are formed from the determinants of $(N-k)\times(N-k)$-dimensional submatrices $A_{[I],[J]}$
that are obtained from $A$ by deleting  $k$ rows with ordered indices $I=[1\leq i_1<i_2<\dots, i_k\leq N]$
and $k$ columns with ordered indices $J=[1\leq j_1<j_2<\dots, j_k\leq N]$.
By definition of the geometric multiplicity, the  matrix $A$ has $N-\gamma_i$ 
linearly dependent rows and columns, so that all minors
\begin{equation}
    M_{[I],[J]}^{(k)}(A)=0, \quad k=0,1,2,\ldots,\gamma_{i}-1 
      \label{eq:mkcondition}
\end{equation}
vanish. This then serves as an algebraic condition for the geometric degeneracy of the eigenvalue $E_i$, which already reads quite analogously to the condition \eqref{eq:ckcondition} for the algebraic degeneracy. 

These considerations can be linked to the modes $\mathcal{B}_k$
by exploiting the partial traces of the determinantal minors, which can be defined as 
\begin{align}
    \mathcal{N}^{(k)}_{i,j}&=\!\!\!
\sum_{[p,q,r...]} \!\!\!
\sigma(i,p,q,r...)\sigma(j,p,q,r...) 
    M^{(k)}_{[i,p,q,r...], [j,p,q,r...]}
    .
    \label{partial2}
\end{align}
These partial traces are obtained by contracting all but one pair of row and column indices, where $\sigma(I)=\pm 1$ is the parity of the permutation that orders the sequence $I$.
With these definitions, we then have the identity
\cite{bid2024uniform}
\begin{equation}
\mathcal{B}_k(A)=\Sigma[\mathcal{N}^{(k+1)}(-A)]^T\Sigma,
\label{eq:bfromn}
\end{equation}
where $\Sigma_{ij}=(-1)^{i}\varepsilon_{ij}$ is the diagonal matrix with alternating signs on the diagonal.

To identify the exact nature of the information contained in the modes $\mathcal{B}_k$, one can evaluate the partial traces $\mathcal{N}^{(k)}$ in the Jordan normal form $J$ of the matrix $A$.
Because of condition  \eqref{eq:ckconditionAPP}, the modes $\mathcal{B}_k$ with $k<\alpha_i$ then take a simple form, in which only the block with the degenerate eigenvalue can have finite entries. Furthermore, within this index range, the recursion relation \eqref{eq:flv} simplifies to $\mathcal{B}_{k-1}=\mathcal{B}_{k}J$. Because of the simple form of the Jordan normal form matrix $J$, we can then directly read off the rank of the matrix $\mathcal{B}_{k}$,
\begin{align}
 \mathrm{rnk}(\mathcal{B}_{\alpha_i-m})
 &=\alpha_i-\sum_{l=1}^{\alpha_i}\mathrm{min}(l,m-1)\beta_i(l)
 \nonumber\\
&= \sum_{l=m}^{\alpha_i}(l-m+1)\beta_i(l),
\end{align}
where we used Eq.~\eqref{partialb} to reexpress $\alpha_i$ in terms of $\beta_i(l)$.
This completes the proof of our main technical result in this work.
The partial multiplicity function $\beta_i(l)$  is then resolved by combining the ranks of modes according to Eq.~\eqref{eq:betalresult}. 
This delivers the geometric multiplicity $\gamma_i$ via Eq.~\eqref{partiala} and the algebraic multiplicity $\alpha_i$ via Eq.~\eqref{partialb}.
Evaluating Eq.~\eqref{eq:betalresult} for $k=\alpha_i-\ell_i$, where the maximal partial multiplicity follows from Eq.~\eqref{eq:rankcondition2},
we furthermore obtain the geometric multiplicity $\beta_i$ of the leading eigenvectors in accordance with Eq.~\eqref{eq:beta}.

\end{document}